\documentclass[3p,sort&compress]{elsarticle}

\usepackage[T1]{fontenc}
\usepackage{lmodern}
\usepackage[english]{babel}\usepackage{csquotes}\usepackage[activate={true,nocompatibility},tracking=true]{microtype}\usepackage{hyphenat}\usepackage{siunitx}
\usepackage{placeins}
\usepackage[begintext=``, endtext='']{quoting}
\SetBlockEnvironment{quoting}
\SetBlockThreshold{1}
\usepackage{url}

\usepackage{lineno,hyperref}
\modulolinenumbers[5]

\usepackage{amsfonts}
\usepackage{amssymb}
\usepackage{amscd}
\usepackage{mathtools}
\usepackage[mathscr]{eucal}
\usepackage{calligra}
\DeclareMathAlphabet{\mathcalligra}{T1}{calligra}{m}{n}
\DeclareFontShape{T1}{calligra}{m}{n}{<->s*[1.1]callig15}{}
\usepackage{bm}
\usepackage[nice]{nicefrac}
\usepackage{bbm}

\usepackage{tikz}
\usepackage{pgf}
\usepackage{pgfplots}
\usepackage{pgfplotstable}
\usepackage{shellesc}
\pgfplotsset{compat=newest}
\usetikzlibrary{spy}
\usetikzlibrary{shapes}
\usetikzlibrary{backgrounds}
\usetikzlibrary{shadows}
\usetikzlibrary{matrix}
\usetikzlibrary{external}
\usepgfplotslibrary{fillbetween}
\tikzset{
	font={\fontsize{10pt}{10}\selectfont}}
\pgfplotscreateplotcyclelist{\myblackwhite}{	thick, solid, \\	thick,densely dotted,\\	thick,dashed,\\							}
\pgfplotsset{every tick label/.append style={font=\scriptsize}}

\usepackage{color}
\definecolor{white}{RGB}{255,255,255}
\definecolor{black}{RGB}{0,0,0}
\definecolor{greenForest}{RGB}{34,139,34}
\definecolor{grayTransparent}{RGB}{83,83,83}
\definecolor{eggYolkYellow}{RGB}{249,208,16}
\definecolor{blueNavy}{RGB}{0,0,128}
\definecolor{redBrick}{RGB}{203,65,84}

\usepackage{comment}
\usepackage{graphicx}
\usepackage{import}
\usepackage[colorinlistoftodos,prependcaption,textsize=small]{todonotes}
\usepackage{cleveref}
\usepackage{autonum}\crefname{equation}{equation}{equations}
\crefname{theorem}{theorem}{theorems}
\crefname{chapter}{chapter}{chapters}
\crefname{figure}{figure}{figures}
\crefname{conject}{conjecture}{conjectures}
\crefname{lemma}{lemma}{lemma}
\crefname{corollary}{corollary}{corollaries}
\crefname{section}{section}{sections}
\crefname{definition}{definition}{definitions}
\crefname{table}{table}{tables}
\crefname{remark}{remark}{remarks}
\crefname{problem}{problem}{problems}
\crefname{proposition}{proposition}{propositions}
\crefname{algocf}{algorithm}{algorithms}

\crefname{equation}{}{}
\usepackage{csvsimple}
\usepackage{booktabs}
\usepackage{environ}
\usepackage{adjustbox}
\usepackage{relsize}
\usepackage[font=small,figurewithin=section]{caption}\usepackage[subrefformat=parens,labelformat=simple]{subcaption}
\usepackage{afterpage}
\usepackage[ruled,vlined,algo2e]{algorithm2e} 
\newtheorem{remark}{Remark}[section]

\let\inf\relax \DeclareMathOperator*\inf{\vphantom{p}inf}
\let\min\relax \DeclareMathOperator*\min{\vphantom{p}min}
\let\max\relax \DeclareMathOperator*\max{\vphantom{p}max}
\let\subset\relax \DeclareMathOperator{\subset}{\subseteq}

\let\hat\widehat

\DeclareMathOperator*{\argmin}{arg\,min}

\newcommand{\rev}[1]{{#1}}

\newcommand{\sspace}{\hspace{0.25pt}}

\newcommand{\rseed}{\mathfrak{\rev{s}}}

 \newcommand{\R}{\mathbb{R}}       \newcommand{\dd}{\,\mathrm{d}}

\newcommand{\bxi}{\bm{\xi}}

\newcommand{\bme}{\bm{e}}

\newcommand{\bmk}{\bm{k}}

\newcommand{\bmn}{\bm{n}}

\newcommand{\bmr}{\bm{r}}

\newcommand{\bmu}{\bm{u}}

\newcommand{\bmx}{\bm{x}}

\newcommand{\bmO}{\bm{O}}

\newcommand{\mcN}{\mathcal{N}}

\newcommand{\scA}{\mathscr{A}}

\newcommand{\bbC}{\mathbb{C}}

\newcommand{\bbE}{\mathbb{E}}

\newcommand{\bbP}{\mathbb{P}}

\newcommand{\bbV}{\mathbb{V}}

\newcommand{\rmE}{\mathrm{E}}

\newcommand{\rmL}{\mathrm{L}}

\newcommand{\rmR}{\mathrm{R}}

\newcommand{\rmU}{\mathrm{U}}

\newcommand{\meansquareerrorchar}{\ensuremath{\operatorname{\emph{mse}}}}
\newcommand{\meansquareerror}[2][2]{
	\ifthenelse{\equal{#2}{}}
	{\ensuremath{\meansquareerrorchar^{#1}}}
	{\ensuremath{\meansquareerrorchar_{\text{#2}}^{#1}}}
}

\newcommand{\meshparchar}{M}
\newcommand{\meshparameter}[1][]{
	\ifthenelse{\equal{#1}{}}
	{\ensuremath{\meshparchar}}
	{\ensuremath{\meshparchar_{#1}}}
}
\newcommand{\numsampleschar}{N}
\newcommand{\numbersamples}[1][]{
	\ifthenelse{\equal{#1}{}}
	{\ensuremath{\numsampleschar}}
	{\ensuremath{\numsampleschar_{#1}}}
}
\newcommand{\meshsizechar}{h}
\newcommand{\meshsize}[1][]{
	\ifthenelse{\equal{#1}{}}
	{\ensuremath{\meshsizechar}}
	{\ensuremath{\meshsizechar_{#1}}}
}

\newcommand{\numberlevels}[1][]{
	\ifthenelse{\equal{#1}{}}
	{\ensuremath{L}}
	{\ensuremath{L_{#1}}}
}

\newcommand{\diffqoichar}{\ensuremath{Y}}
\newcommand{\differenceqoi}[1][]{
	\ifthenelse{\equal{#1}{}}
	{\ensuremath{\diffqoichar}}
	{\ensuremath{\diffqoichar_{#1}}}
}

\newcommand{\timechar}{T}
\newcommand{\totaltime}[1][]{
	\ifthenelse{\equal{#1}{}}
	{\ensuremath{\timechar}}
	{\ensuremath{\timechar_{#1}}}
}

\newcommand{\CDFchar}{\ensuremath{\Phi}}
\newcommand{\CDFsymbol}[1][]{
	\ifthenelse{\equal{#1}{}}
	{\ensuremath{\CDFchar}}
	{\ensuremath{\CDFchar(#1)}}
}
\DeclareMathOperator{\expectedvaluechar}{\bbE}
\newcommand{\expectedvalue}[1][]{
	\ifthenelse{\equal{#1}{}}
	{\ensuremath{\expectedvaluechar}}
	{\ensuremath{\expectedvaluechar[#1]}}
}
\newcommand{\expectedvalueMC}[1][]{
	\ifthenelse{\equal{#1}{}}
	{\ensuremath{\expectedvaluechar^{\text{\MC}}}}
	{\ensuremath{\expectedvaluechar^{\text{\MC}}[#1]}}
}
\newcommand{\expectedvalueMLMC}[1][]{
	\ifthenelse{\equal{#1}{}}
	{\ensuremath{\expectedvaluechar^{\text{\MLMC}}}}
	{\ensuremath{\expectedvaluechar^{\text{\MLMC}}[#1]}}
}
\DeclareMathOperator{\variancechar}{\bbV}
\newcommand{\variancevalue}[1][]{
	\ifthenelse{\equal{#1}{}}
	{\ensuremath{\variancechar}}
	{\ensuremath{\variancechar[#1]}}
}
\newcommand{\variancevalueMC}[1][]{
	\ifthenelse{\equal{#1}{}}
	{\ensuremath{\variancechar^{\text{\MC}}}}
	{\ensuremath{\variancechar^{\text{\MC}}[#1]}}
}
\DeclareMathOperator{\probabilitychar}{\bbP}
\newcommand{\probability}[1][]{
	\ifthenelse{\equal{#1}{}}
	{\ensuremath{\probabilitychar}}
	{\ensuremath{\probabilitychar[#1]}}
}

\newcommand{\samplevarchar}{\ensuremath{\bar{\sigma}}}
\newcommand{\samplevar}[1][]{
	\ifthenelse{\equal{#1}{}}
	{\ensuremath{\samplevarchar}}
	{\ensuremath{\samplevarchar_{#1}}}
}

\newcommand{\centralmoment}[1][]{
	\ifthenelse{\equal{#1}{}}
	{\ensuremath{\mu}}
	{\ensuremath{\mu_{#1}}}
}
\newcommand{\powersum}[1][]{
	\ifthenelse{\equal{#1}{}}
	{\ensuremath{S}}
	{\ensuremath{S_{#1}}}
}
\newcommand{\hstatistics}[1][]{
	\ifthenelse{\equal{#1}{}}
	{\ensuremath{h}}
	{\ensuremath{h_{#1}}}
}
\DeclareMathOperator{\computationalcostchar}{\bbC}
\newcommand{\computationalcost}[1][]{
	\ifthenelse{\equal{#1}{}}
	{\ensuremath{\computationalcostchar}}
	{\ensuremath{\computationalcostchar[#1]}}
}

\newcommand{\domainchar}{\Omega}
\newcommand{\domain}[1][]{
	\ifthenelse{\equal{#1}{}}
	{\ensuremath{\domainchar}}
	{\ensuremath{\domainchar_{#1}}}
}

\newcommand{\solchar}{u}
\newcommand{\sol}[1][]{
	\ifthenelse{\equal{#1}{}}
	{\ensuremath{\solchar}}
	{\ensuremath{\solchar_{#1}}}
}
\newcommand{\randvarchar}{w}
\newcommand{\randomvariable}[1][]{
	\ifthenelse{\equal{#1}{}}
	{\ensuremath{\randvarchar}}
	{\ensuremath{\randvarchar^{#1}}}
}
\newcommand{\qoi}[1][]{
	\ifthenelse{\equal{#1}{}}
	{\ensuremath{\qoitext}}
	{\ensuremath{\qoitext_{#1}}}
}

\newcommand{\convergencetolerance}[1][]{
	\ifthenelse{\equal{#1}{}}
	{\ensuremath{\varepsilon}}
	{\ensuremath{\varepsilon_{#1}}}
}
\newcommand{\convergenceconfidence}[1][]{
	\ifthenelse{\equal{#1}{}}
	{\ensuremath{\phi}}
	{\ensuremath{\phi_{#1}}}
}

\newcommand{\MC}{MC}
\newcommand{\MLMC}{MLMC}

\newcommand{\qoitext}{Q}

\journal{Computer Methods in Applied Mechanics and Engineering}

\begin{document}

\begin{frontmatter}

\title{
		Risk-averse design of tall buildings for uncertain wind conditions}
	
\author[1]{Anoop~Kodakkal} \ead{anoop.kodakkal@tum.de}

\author[2]{Brendan~Keith}
\ead{brendan\_keith@brown.edu}
\author[3]{Ustim~Khristenko}
\ead{khristen@ma.tum.de}
\author[1]{Andreas~Apostolatos}
\ead{Andreas.Apostolatos@tum.de}
\author[1]{Kai-Uwe~Bletzinger}
\ead{kub@tum.de}
\author[3]{Barbara~Wohlmuth}
\ead{wohlmuth@ma.tum.de}
\author[4]{Roland~W\"uchner}
\ead{r.wuechner@tu-braunschweig.de}

\address[1]{Chair of Structural Analysis, TUM School of Engineering and Design, Technical University of Munich, Arcisstr. 21, 80333 Munich, Germany}
\address[2]{Divison of Applied Mathematics, Brown University, 170 Hope St, Providence, RI, 02912, USA}
\address[3]{Chair of Numerical Mathematics, Technical University of Munich, Arcisstr. 21, 80333 Munich, Germany}
\address[4]{Institute of Structural Analysis, Technische Universit\"at Braunschweig, Beethovenstr. 51, 38106 Braunschweig, Germany}

\begin{abstract}
Reducing the intensity of wind excitation via aerodynamic shape modification is a major strategy to mitigate the reaction forces on supertall buildings, reduce construction and maintenance costs, and improve the comfort of future occupants. To this end, computational fluid dynamics (CFD) combined with state-of-the-art stochastic optimization algorithms is more promising than the trial and error approach adopted by the industry. The present study proposes and investigates a novel approach to risk-averse shape optimization of tall building structures that incorporates site-specific uncertainties in the wind velocity, terrain conditions, and wind flow direction. A body-fitted finite element approximation is used for the CFD with different wind directions incorporated by re-meshing the fluid domain. The bending moment at the base of the building is minimized, resulting in a building with reduced cost, material,  and hence, a reduced carbon footprint. Both risk-neutral (mean value) and risk-averse optimization of the twist and tapering of a representative building are presented under uncertain inflow wind conditions that have been calibrated to fit freely-available site-specific data from Basel, Switzerland. The risk-averse strategy uses the conditional value-at-risk to optimize for the low-probability high-consequence events appearing in the worst 10\% of loading conditions. Adaptive sampling is used to accelerate the gradient-based stochastic optimization pipeline. The adaptive method is easy to implement and particularly helpful for compute-intensive simulations because the number of gradient samples grows only as the optimal design algorithm converges. The performance of the final risk-averse building geometry is exceptionally favorable when compared to the risk-neutral optimized geometry, thus, demonstrating the effectiveness of the risk-averse design approach in computational wind engineering.
\end{abstract}


\end{frontmatter}

{\bf Dedication:}
The authors dedicate this work to Dr. J. Tinsley Oden as a token of everlasting respect and admiration for his monumental contributions to science and education.
J.T. Oden is an indomitable pioneer and true game-changer in Computational Science and Engineering.
His vision and leading role have provided the archetype of interdisciplinary research.
Throughout his prodigious career, J.T. Oden has shaped and promoted predictive science to make it a discipline built upon deep concepts in mathematics, computer sciences, physics, and mechanics.
We remain forever in debt, awe, and gratitude to this great man.

\section{Introduction} \label{sec:introduction}
Building structures are becoming taller and more flexible due to improved design methods, new materials, and novel construction technologies. The structural design of these supertall buildings is primarily driven by lateral loads such as wind forces. Referring only to design codes~\cite{ASCE7-982002, EN1991-4} for these supertall buildings is undesirable. Indeed, designers need to precisely know the wind forces acting on a tall building as such buildings generally have non-trivial cross-sections and geometry, resulting in complex wind flow patterns. For this reason, it is a standard practice to use wind tunnel tests to determine the wind behavior around tall buildings. Nevertheless, over the past few decades, computational wind engineering (CWE) has matured enough to accurately predict the pressure field and other wind effects on civil structures~\cite{Blocken2014}. Blocken~\cite{Blocken2014} reiterates the need for high-quality and reliable computational fluid dynamics (CFD) simulations in practical applications, and, in recent years, CWE has become much more widely used to analyze and design various structures and tall buildings~\cite{pentek18, andre_michael_jweia17}.

Geometric modifications are very effective at controlling wind loads on tall buildings \cite{tanaka2013aerodynamic}; the interested reader is referred to~\cite{AsghariMooneghi2016} for a summary of various beneficent global and local geometric modifications.
Of note is that most previous engineering studies have relied on wind tunnel tests to select geometric modifications.
For instance, the effect of openings in a tall rectangular building was studied with a boundary layer wind tunnel~\cite{DUTTON1990739}.
Here, it is reported that the through-building opening effectively reduces the across wind excitation. In~\cite{Carassale2014}, the effect of corner roundness on a square cross-section is also evaluated using wind tunnel tests.

\begin{figure}
    \centering
    \begin{subfigure}{.3\textwidth}
        \centering
        \includegraphics[height=1.2\linewidth]{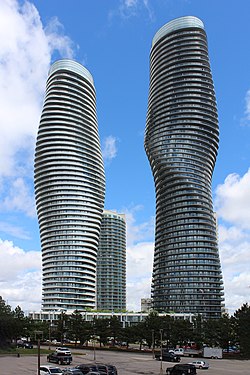}
        \caption{}
        \label{fig:absolute_world}
    \end{subfigure}    \begin{subfigure}{.3\textwidth}
        \centering
        \includegraphics[height=1.2\linewidth]{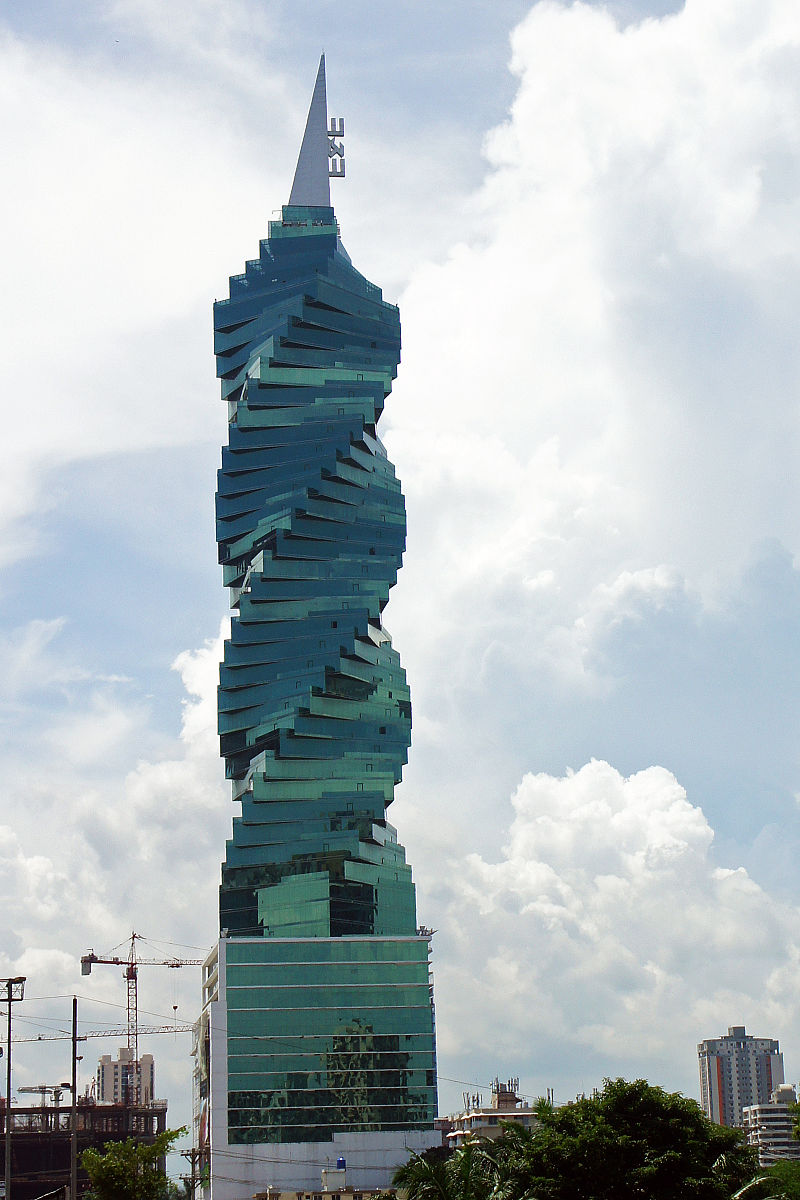}
        \caption{}
        \label{fig:panamafandf}
    \end{subfigure}
    \begin{subfigure}{.3\textwidth}
        \centering
        \includegraphics[height=1.2\linewidth]{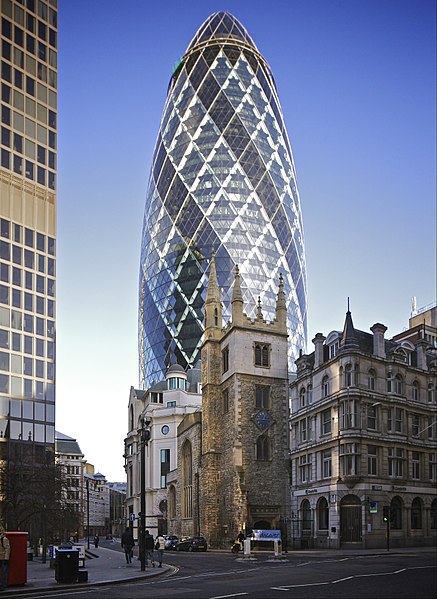}
        \caption{}
        \label{fig:St_Mary_Axe}
    \end{subfigure}
\caption{Modern tall buildings with non-trivial shapes: (a) Absolute World Towers, Mississauga, Canada with elliptical cross section and twist along the height \cite{bahga_2018}; (b) F\&F tower, Panama with twist along the height \cite{fandf}; (c) 30 St Mary Axe tower, with circular dome at the roof, London, United Kingdom \cite{St_mary_axe_tower}. }
\label{fig:twisted_tower}
\end{figure}

The primary drawback of using wind tunnel tests to determine wind mitigation strategies is that specific expertise, costly experiments, repeated model building, and repeated testing is required to determine optimal geometries. Computational methods provide an environment to accelerate this process as well as incorporate site-specific uncertainties that match experimental measurements that are not easily reproducible in a wind tunnel environment. One candidate optimization framework is presented in~\cite{Elshaer2017}. In that work, both (local) corner modifications and (global) shape modifications are considered. Nevertheless,~\cite{Elshaer2017} does not consider any environmental uncertainties, and it relies on a non-rigorous neural network-based genetic algorithm to perform the optimization. 

Wind exhibits uncertainty. This fact is perhaps most evident from the fluctuating component of the wind velocity field. However, the mean wind velocity and the mean wind direction are also inherently random variables. Furthermore, the mean wind profile manifests its own uncertainties coming from the effects of the local terrain. To arrive at a robust and reliable design, one needs to take many uncertainties into consideration during analysis \cite{Davenport2002}. The quantification of these uncertainties in wind turbine response is well-studied in the literature \cite{Lackner2007, VanDenBos2018}.
By comparison, uncertainty quantification for other tall structures suggests room for improvement, especially in the building design process.
To the best of our knowledge, stochastic optimization of geometric building parameters under the uncertainties that arise from site-specific wind conditions has not been investigated in the literature.

A complete framework for optimizing building geometries under the uncertainties of the incoming wind and site conditions is introduced here. More specifically, global geometric parameters are optimized to reduce the norm of the bending moment at the base of the building that results from incoming wind flows. Inspired by the Absolute World Towers and other twisted and tapered supertall buildings (cf. \Cref{fig:twisted_tower}), our numerical experiments focus solely on tapering and twisting.
Nevertheless, our methods are general enough to be also used for alternative geometric modifications such as adjusting the size and shape of an opening and adjusting the size and location of ``stepping'' features reported in ~\cite{Irwin2009}.
Given the uncertain nature of the loading patterns resulting from the incoming wind, various stochastic optimization problems may be formulated to achieve our ends.
For instance, we may consider a ``risk-neutral'' problem formulation; wherein we optimize for only the mean value of our quantity of interest (i.e., the bending moment).
Alternatively, we may consider robust optimization based on the mean and standard deviation \cite{Elshaer2017b} or reliability-based optimization \cite{Hu2016}.

In this work, we choose to focus on ``risk-averse'' optimization by the conditional value-at-risk ($\mathrm{CVaR}$) \cite{rockafellar2000optimization,kouri2016risk,kouri2018optimization}.
Although $\mathrm{CVaR}$ is a measure of risk that initially appeared in finance \cite{rockafellar2000optimization}, it has been found in recent years to be useful for engineering design \cite{rockafellar2015engineering,kouri2016risk,kouri2018optimization,beiser2020adaptive}. In particular, as argued in \cite{rockafellar2015engineering}, it is mathematically superior to robust and reliability-based optimization problem formulations when controlling for low-probability events \cite{rockafellar2010buffered,rockafellar2013fundamental,rockafellar2015engineering}. Risk-neutral and risk-averse optimization are compared in the present study. 
All statistics in our workflow are estimated via the Monte Carlo method.
Furthermore, we use a stochastic gradient descent method \cite{wright2022optimization} for numerical optimization and a novel adaptive sampling strategy \cite{byrd2012sample,bollapragada2018adaptive,xie2020constrained,beiser2020adaptive} to reduce the optimization cost.

It is known that the incidence angle of the wind and its speed are correlated random variables.
In this work, these and other site-specific parameters in the simulation environment are calibrated using freely-available historical data.

Under uncertainty, site-specific optimal design is associated with unique challenges, including increased computational cost. However, this effort and cost are justified if it is considered that tall buildings are one-time built structures with monumental expenses if they fail. Moreover, the building foundation and other features are extremely difficult to retrofit for future changes in the building due to, e.g., unexpected loads or human error during design. Considering uncertainty in the design decision-making will help minimize risks from these scenarios and reduce the cost of retrofitting due to rare events.

The rest of the paper is structured as follows:
\Cref{sec:accounting_for_uncertainties} describes the physical uncertainties in the simulation environment and the mathematical models we have used to account for them. \Cref{sec:building_designs} introduces the computer-aided building design space and the finite element model we used for the computational fluid dynamics simulation. In \Cref{sec:chance_constraints}, we discuss certain principles of risk measurement, the conditional value-at-risk, and the corresponding design optimization problems. \Cref{sec:optimization_algorithm} describes the adaptive stochastic optimization algorithm and the stochastic optimization workflow. \Cref{sec:results_and_candidate_designs} shows the results of our numerical studies and the two different optimization strategies. Finally, in~\Cref{sec:conclusions}, we present our conclusions.

\section{Accounting for uncertainties} \label{sec:accounting_for_uncertainties}

In this section, we introduce two models for location-specific uncertainty quantification in tall building wind engineering.
As is typical in the wind engineering community, we model the natural wind effects in the atmospheric boundary layer by decomposing the incoming velocity field, $\bmu = \overline{\bmu} + \bmu^\prime$, into its steady \emph{mean profile} $\overline{\bmu}$ and its unsteady \emph{turbulent fluctuations} $\bmu^\prime$.
The mean wind profile $\overline{\bmu}$ is a contribution to the overall velocity field $\bmu$ that changes gradually over the span of several hours or days \cite{kareem2015advanced}.
For this reason, it is generally considered constant with respect to the scale of most numerical simulations.
On the other hand, the turbulent fluctuations $\bmu^\prime$ introduce short term wind gusts with a time span of seconds or minutes.
Due to the different time scales above, we choose to use separate and independent statistical models for $\overline{\bmu}$ and $\bmu^\prime$.
Once combined, these two models allow us to represent the most influential uncertainties that drive the simulation.
We begin with a data-driven statistical model for the mean profile $\overline{\bmu}$ and complement it with a standardized model for the turbulent fluctuations $\bmu^\prime$.

\subsection{Wind modeling: The mean profile} \label{sub:incoming_wind}

Most high-rise buildings reside entirely within the atmospheric boundary layer (ABL); a layer of Earth's atmosphere, extending vertically from its surface, that is characterized by constant shear stress in the vertical direction \cite{kaimal1994atmospheric}.
This region is generally recognized to be \emph{neutrally stable} at high wind speeds.
That is, the buoyancy forces due to temperature gradients are negligible in comparison to surface-driven friction forces.

Ground friction is dominated by pressure drag; i.e., the force generated by pressure differences near the surface and caused by wind flowing across obstacles.
Depending on the local terrain, the variety of obstacles affecting ground friction can change vastly.
For instance, consider that different friction forces will arise from flow across grass, forests, open water, or urban canopies.
In this study, we assume that the local terrain type can be sufficiently characterized by a single \emph{roughness length} parameter $z_0>0$.
Intervals of validity for this parameter, for various terrain categories, can be found in numerous engineering codebooks; see, e.g., \cite{jcss2001probabilistic}.
In addition, our mean profile model incorporates the \emph{incidence wind angle} $\theta$ and the \emph{friction velocity} $u_\ast$. The frictional velocity can be derived from the shear stress on the ground $\tau_0$ by the simple formula $\tau_0 = \rho u_\ast^2$ \cite{mann1998wind}.

\begin{figure}
	\centering
	\includegraphics[width=0.93\textwidth]{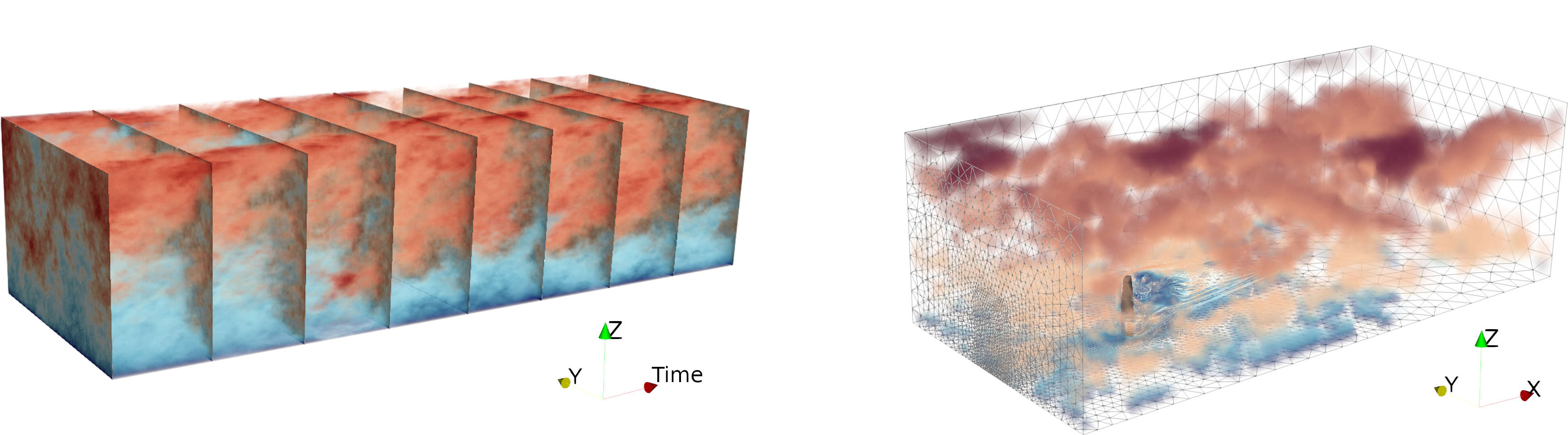}
	\raisebox{0.3\height}{\includegraphics[width=0.06\textwidth]{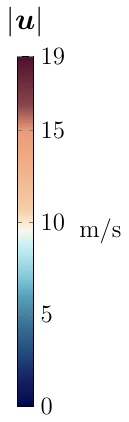}}
\caption{\label{fig:WindSnapshots} Synthetic turbulent wind fluctuations $\bmu^\prime(t,y,z)$ are generated as a 3D block in the $(t,y,z)$-space (left).
	Snapshots corresponding to different times~$t$ are used to impose the inlet boundary velocity $\bmu(t,x=0, y, z)= \overline{\bmu}(z) + \bmu^\prime(t,y,z)$ for the CFD computational domain (right).}
\end{figure}

Let $(x,y,z)$ denote Cartesian coordinates in the region of the ABL surrounding the building; see, e.g., the right-hand side of~\Cref{fig:WindSnapshots}.
Furthermore, let the unit normal vector $\bme(\theta)=(\cos(\theta),\sin(\theta),0)\in \R^3$ denote the mean wind direction.
Under the assumptions of neutral stability and homogeneous roughness, the mean velocity $\overline{\bmu} = \overline{\bmu}(z)$ can be modeled by the following quasi-logarithmic profile \cite{mann1998wind}:
\begin{equation}
\label{eq:MeanProfile}
		\overline{\bmu}(z) =
	\begin{cases}
		\frac{1}{\kappa}\Big(u_\ast\ln\Big(\frac{z}{z_0}\Big) + 34.5f z\Big)\sspace\bme(\theta)
		&
		\text{if~} z \geq z_{\min}
		,
		\\
		\overline{\bmu}(z_{\min})\sspace\bme(\theta)
		& \text{otherwise,}
	\end{cases}
		\end{equation}
where $\kappa \approx 0.41$ is the von Karm\'{a}n constant, z$_{min} > z_0$ is the minimum height defined in \cite{EN1991-4}, and $f =10 ^{-4}$ is the Coriolis parameter.

Each of the parameters $u_\ast$, $\theta$, and $z_0$ in~\cref{eq:MeanProfile} are random variables.
It is often assumed that the friction velocity, averaged over all angles $\theta$, obeys a Weibull distribution, $\mathrm{Wieb}(\lambda,k)$, with \emph{scale} $\lambda$ and \emph{shape} $k$ \cite{kareem2015advanced}.
Likewise, statistical models for the wind angle $\theta$ are often constructed using a mixture of von Mises distributions \cite{carnicero2013non,garcia2013exploring}.
Models for the \rev{roughness length} $z_0$ are only recently studied in detail \cite{kent2018surface,kent2018aerodynamic}.
In this work, we assume that $z_0$ follows a uniform distribution, $z_0\sim \mathrm{Unif}(z_\rmL,z_\rmU)$, where $z_\rmL$ and $z_\rmU$ are positive constants inferred from \cite{jcss2001probabilistic}.

\begin{figure}
	\centering
	\begin{subfigure}[c]{0.39\textwidth}
	\caption*{\normalsize Field data}
	        \centering
	        \includegraphics[clip=true,trim=2.1cm 0cm 2cm 0cm,width=\textwidth]{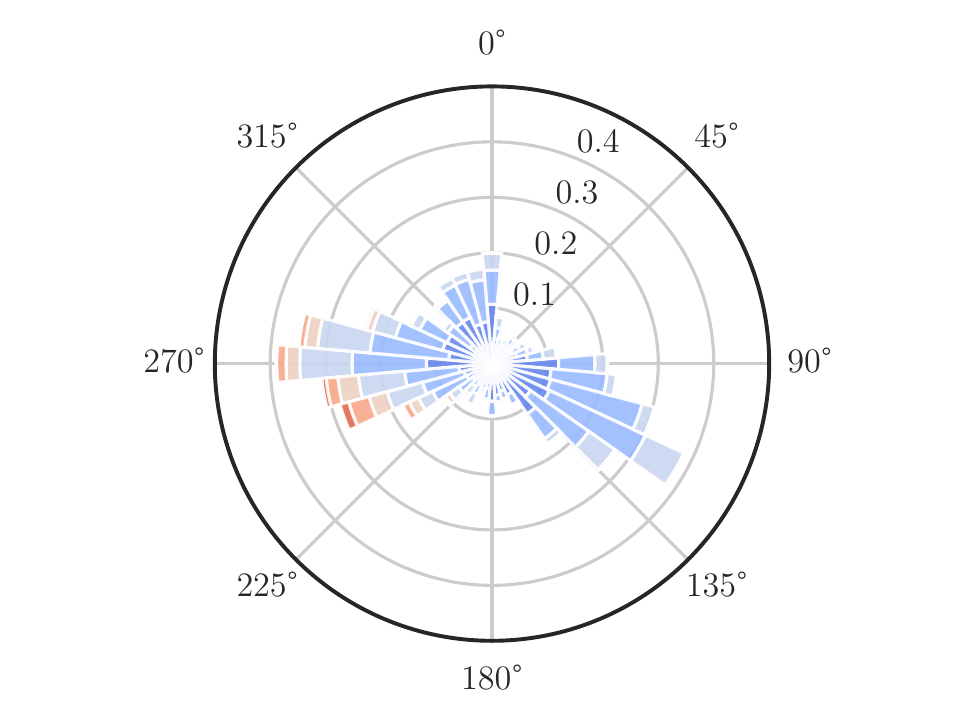}
		\end{subfigure}
		\begin{subfigure}[c]{0.39\textwidth}
	\caption*{\normalsize Synthetic data}
	        \centering
	        \includegraphics[clip=true,trim=2.1cm 0cm 2cm 0cm,width=\textwidth]{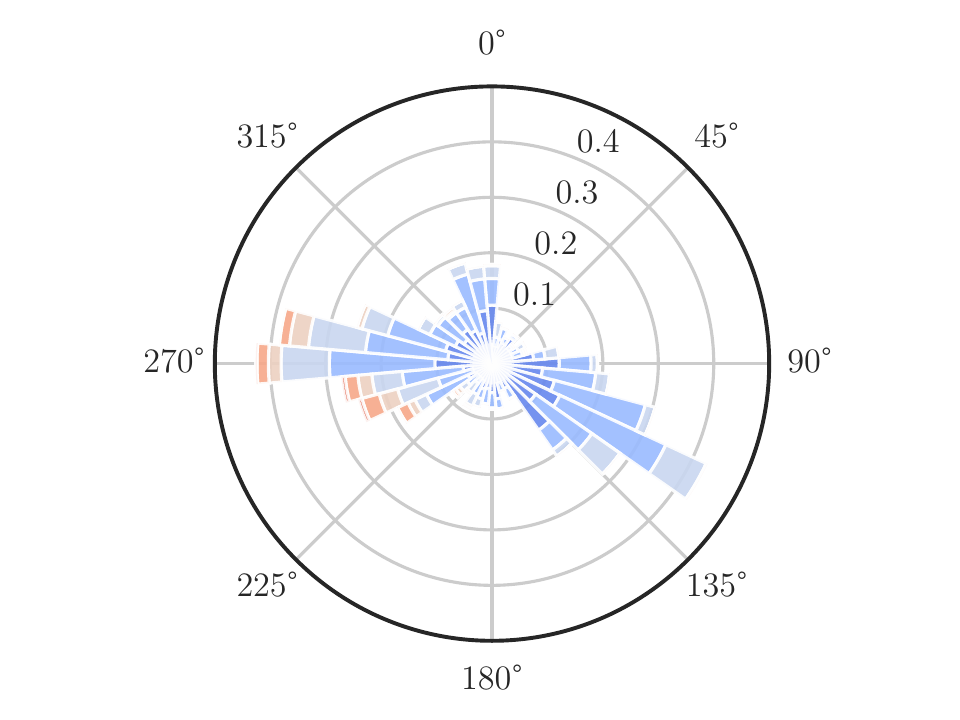}
		\end{subfigure}
	~
	\begin{subfigure}[c]{0.16\textwidth}
	\caption*{}
	        \centering
	        \includegraphics[clip=true,trim=0.1cm 0.1cm 6.6cm 0.1cm,width=\textwidth]{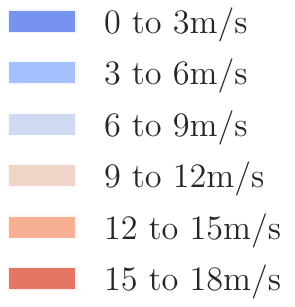}
		\end{subfigure}
\caption{\label{fig:WindRoses} Wind roses representing the mean wind speed $\bar{u}$ and incidence angle $\theta$ parameters in the mean profile model. The field data was collected at $z=80\mathrm{m}$ in Basel, Switzerland between 2010-01-01 and 2015-12-31.
}
\end{figure}

If $u_\ast$, $\theta$, and $z_0$ are independent random variables, then the assumptions in the previous paragraph would be enough to form a complete, parameterized statistical model for $\overline{\bmu} = \overline{\bmu}(u_\ast,\theta,z_0)$.
However, in most environments, independence of these random variables is a very poor assumption because of nearby geographic features or persistent weather patterns.
For illustration, consider the wind rose on the left-hand side of~\Cref{fig:WindRoses}, which indicates a strong dependence between the mean wind speed $\overline{u} := |\overline{\bmu}|$ and angle $\theta$ in field measurements at $z=80\mathrm{m}$ in Basel, Switzerland. 

In this work, we assume that $z_0$ is independent of both $u_\ast$ and $\theta$ because the data necessary to infer this dependence is not widely available.
On the other hand, data respresenting the distribution of $u_\ast$ and $\theta$ is widely available.
We choose to use an established class of bivariate copulas \cite{carnicero2013non,garcia2013exploring} to model the dependence between these two random variables.
Our particular model is described in \ref{sec:copula_based_model_for_wind_speed_and_direction}.
To calibrate this stastical model, we use a data set of mean wind speeds $\overline{u}$ and directions $\theta$ collected at $z=80\mathrm{m}$ above ground in Basel, Switzerland, from 2010-01-01 to 2015-12-31.\footnote{Available for free at \url{https://www.meteoblue.com}.}\footnote{Note that $u_\ast = (\kappa\overline{u} - 34.5fz) / \ln(z/z_0)$ by~\cref{eq:MeanProfile}. Therefore, a statistical model for $\theta$ and $\overline{u}$, at a fixed height $z$, can also be used as a statistical model for $\theta$ and $u_\ast$.\label{foot:ubar}}
\Cref{fig:WindRoses} depicts the field data alongside synthetic data resulting from our copula-based model.

\subsection{Wind modeling: Synthetic turbulent fluctuations} \label{sub:synthetic_wind_generation}

Modeling turbulent fluctuations with physical wind gust statistics can be more challenging than modeling the mean profile.
Numerous techniques have been proposed in the engineering literature to tackle this problem, and we refer the interested reader to \cite{tabor2010inlet,sanderse2011review,wu2017inflow,keith2021learning} for an overview.
In this study, we choose to model the velocity fluctuations using the atmospheric boundary layer turbulence model proposed in \cite{mann1994spatial,mann1998wind}; hereafter, referred to as the \emph{Mann model}.
The Mann model is a widely used spectral model for synthetic wind generation in the atmospheric boundary layer; see, e.g., \cite{tc882005iec,Michalski2011,Andre2015,keith2021fractional,dong2021turbulence}.

The Mann model can be obtained from a mass-conserving linearization of the Navier--Stokes equations under uniform-shear stress assumption at high Reynolds numbers.
This linearization induces a covariance tensor $R_{ij}(\bmr) = \overline{u_i^\prime(\cdot)\sspace u_j^\prime(\cdot + \bmr)}$ for fully-developed homogeneous turbulence that can be combined with a separate wavenumber-dependent eddy lifetime model in order fit field data \cite{mann1994spatial}.

In the Mann model, the random velocity field is defined through the Fourier transform of the covariance tensor~$R(\bmr)$; namely, the so-called \emph{velocity-spectrum tensor} $\Phi_{ij}(\bmk) = \frac{1}{(2\pi)^3}\int_{\R^3} \mathrm{e}^{-\mathrm{i}\bmk\cdot\bmr}R_{ij}(\bmr)\dd \bmr$.
As such, the synthetic turbulent fluctuations $\bmu^\prime = (u_1^\prime,u_2^\prime,u_3^\prime)$ are defined by the inverse Fourier transform,
\begin{equation}
\label{eq:MannFourierSeries}
u_i^\prime(\bmx)
=
\int_{\R^3} \mathrm{e}^{\mathrm{i}\bmk\cdot\bmx}\sum_{j=1}^{3}C_{ij}(\bmk)\xi_j(\bmk)\dd \bmk
\,,
\end{equation}
where the positive-definite second-order tensor $C(\bmk)$ is defined by the velocity-spectrum~$\Phi$ via $C(\bmk) C^{\dagger}(\bmk) = \Phi(\bmk)$, with $\dagger$ denoting complex conjugation of a matrix.
Above, $\xi_j, j=1,2,3$, denotes Gaussian noise in~$\R^3$, i.e., each $\xi_j(\bmk)\sim \mcN(0,1)$ is independent and identically distributed (iid) complex standard normal random variable, and the integral in~\eqref{eq:MannFourierSeries} is understood in the Fourier--Stieltjes sense.
On a uniform tensorial grid in~$\R^3$, with periodic boundary conditions, this integral can be approximated using the Fast Fourier transform; see, e.g., \cite{mann1998wind} for discretization details.
\rev{A spatially-varying three-dimensional block of wind can be interpreted as a time-varying velocity field through Taylor's frozen turbulence hypothesis \cite{Taylor}.  This time-varying wind field can then be mapped to the inlet of the computational domain of a CFD simulation; cf.~\Cref{fig:WindSnapshots}. 
We refer to \cite{mann1998wind,tc882005iec,keith2021learning} for the explicit formula for $\Phi(\bmk)$ and further details of the implementation of the wind generation process based on~\cref{eq:MannFourierSeries}.}
\rev{Note that generating $\xi_j$ requires a pseudorandom number generator.
In order to guarantee distinct random fields within independent parallel executions, the  seeds~$\rseed$ of the corresponding pseudorandom number generators are first drawn from the uniform distribution in $[\rseed_{\rmL},\rseed_{\rmU}]$, up to machine precision, and are then fed to the process scheduler.}

\begin{table}
	\caption{Random variables in the velocity field model $\bmu = \overline{\bmu} + \bmu^\prime$.}
	\centering{}
	\begin{tabular}{llll}
	  \toprule
	  Random variable  & {Distribution} & {Parameters} & \rev{Unit}   \\
	  \midrule
	  Friction velocity ($u_\ast$)   & Weibull           & See \ref{sec:copula_based_model_for_wind_speed_and_direction}.  & \rev{m/s}                 \\
	  Wind direction ($\theta$)       & von Mises          & See \ref{sec:copula_based_model_for_wind_speed_and_direction}. &      \\
	  Roughness length ($z_0$)       & Uniform            & $z_{\rmL} = 0.01$, $z_{\rmU} = 0.1$  & \rev{m}      \\
	  Random seed for turbulence fluctuations ($\rseed$) & Uniform         & $\rseed_{\rmL} = 0$, $\rseed_{\rmU} = 1$ &      \\
	  	  \bottomrule
	\end{tabular}
	\label{table:uncertainties}
  \end{table}

\begin{figure}
	\centering
	\includegraphics[width=\textwidth]{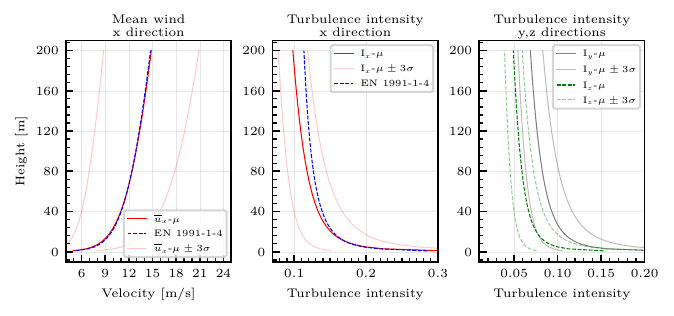}
\caption{\label{fig:Windprofile} Velocity profile (left), turbulence intensity in x direction vs. the height $z$ of the fluid domain (middle), and turbulence intensity in y,z directions vs. the height $z$ of the fluid domain (right).
}
\end{figure}

Details of the random variables in this section are summarized in \Cref{table:uncertainties}. 
\Cref{fig:Windprofile} shows the magnitude of the mean wind profile with respect to the height variable $z$. The turbulence intensity at a height $z$ along the direction $e_i$ is defined as the temporal standard deviation of the turbulence ($\sigma_{i}$) divided by the mean wind speed ($\overline{u}(z)$), namely
\begin{equation}
 	\mathrm{I}_{i}(z) = \frac{\sigma_{i}}{\overline{u}(z)}
 	.
	\label{eq:TurbulenceIntensity}
\end{equation}

\rev{To validate} the wind generation procedure, the turbulence intensity along the height of the domain is compared with the turbulence intensity suggested by Eurocode EN-1991-1-4 \cite{EN1991-4} at the nominal value of the roughness length  $z_0=0.05$\rev{m} \cite{EN1991-4}.\footnote{The roughness length $z_0=0.05$\rev{m} is set for terrain category II, which corresponds to areas with low vegetation and isolated obstacles in joint committee on structural safely \cite{jcss2001probabilistic}.} It can be seen from the left and middle plots in~\Cref{fig:Windprofile} both profiles are in good agreement. The standard deviation $\sigma_{\bmu}$ is computed from 350 independent samples of the wind velocity model $\bmu = \overline{\bmu} + \bmu^\prime$ with the random variables given in~\Cref{table:uncertainties}.

\section{CFD and building designs} \label{sec:building_designs}
The complex wind loads and flow patterns around tall buildings are greatly influenced by geometric features.
For this reason, high fidelity computational fluid dynamics (CFD) are typically required to accurately analyze the effects of wind flow around tall buildings.
In this section, we describe how we have performed the required CFD calculations.
Our implementation uses the open source finite element analysis software Kratos Multiphysics \cite{dadvand2010object,Dadvand2013} and the MMG domain meshing tool \cite{mmg}.

\subsection{Accurate geometries and simulations} \label{sub:details_numerical_model}
\begin{figure}
	\centering
	\begin{minipage}{0.56\linewidth}
		\includegraphics[width=\linewidth]{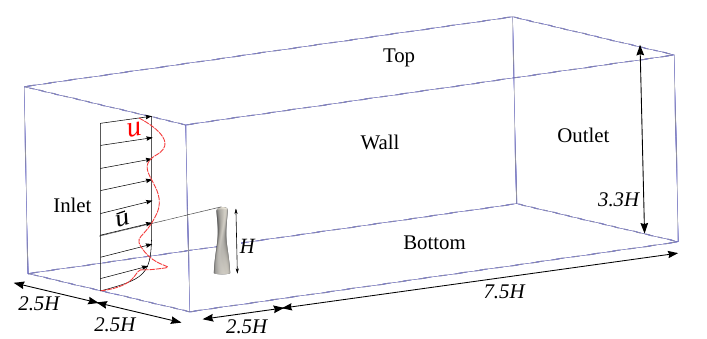}
	\end{minipage}
	~
	\begin{minipage}{0.42\linewidth}
		\footnotesize
		\begin{tabular}{ll}
			\toprule
			CFD domain             & Boundary condition  \\
			\midrule
			Inlet                  & Prescribed velocity (logarithmic) \\
			Outlet                 & Prescribed Pressure (0Pa) \\
			Bottom                 & No-slip condition \\
			Building               & No-slip condition \\
			Top                    & Slip condition \\
			Wall                   & Slip condition \\
			\bottomrule
	  \end{tabular}
	\end{minipage}
	\caption{Details of the flow domain $D$ and boundary conditions. The length scale is the building height $H$.}
	\label{fig:CfdDomain}
		\end{figure}

The wind flow around a tall building is modeled by the incompressible Navier--Stokes equations,
\begin{equation}
	\begin{aligned}
	  \frac{\partial \bmu}{\partial t} + \bmu\cdot\nabla\bmu - \nu \Delta\bmu + \nabla p &= \mathbf{f}
	  \ &&\text{in } D\setminus B\,, \text{ for all } t \in [T_0,T_1] \,,
	  \\
	  \nabla\cdot \bmu
	  &=
	  0
	  \ &&\text{in } D\setminus B\,, \text{ for all } t \in [T_0,T_1] \,,
	  \label{NSEquation}
	\end{aligned}
  \end{equation}
where $\bmu$ is the velocity, $ p $ is the pressure, $ \nu $ is the kinematic viscosity, $ \mathbf{f}$ is the force of gravity, $D$ is a fixed channel domain, $B$ is the building, possibly changing during the optimization process, $D\setminus B$ is the computational domain, and $[T_0,T_1]$ is the time interval.
The boundary conditions and computational domain are depicted in \Cref{fig:CfdDomain}.
We note that, the domain has an inflow region large enough to develop a flow field from the inlet and a sufficiently large outflow region to develop and dissipate vortices.
Note that the length scale of the domain is determined by the building height $H$.

\begin{figure}
	\centering
	\includegraphics[trim=0cm 1.2cm 0cm 0cm, clip=true,width=1.0\textwidth]{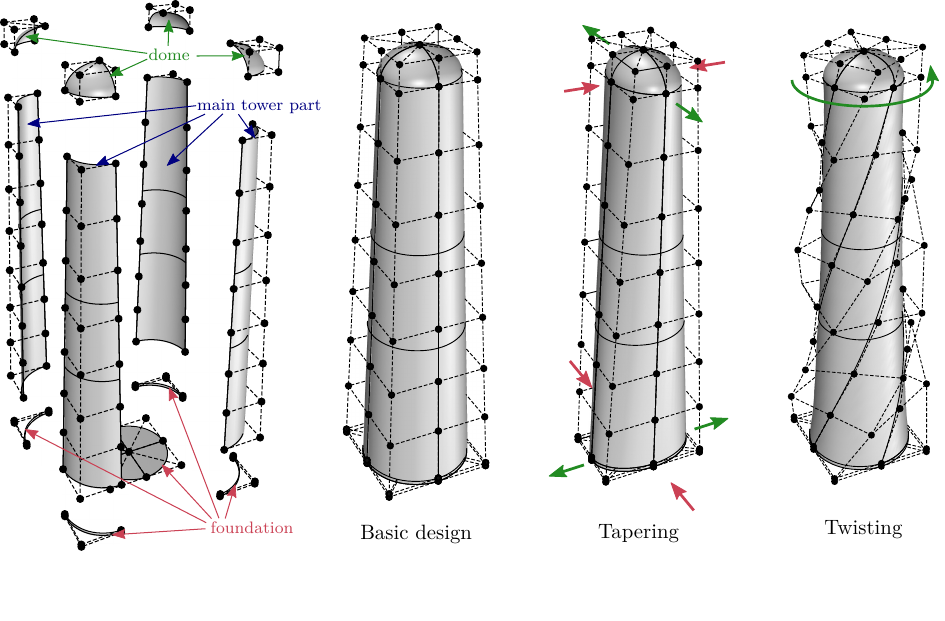}
	\caption{Computer-aided design model of the building $B$. The design space is parameterized by the amount of tapering and twisting.}
		\label{fig:tower_setting}\end{figure}

\Cref{fig:tower_setting} shows the various tower parts and how the design modifications are incorporated into the current study. The basic design is an elliptical cross-section with a dome at the top. The two global geometric modifications allowed are tapering and twisting. The shape of the building $B$ is determined by these parametric modifications, and its orientation is determined by the direction of the incoming wind.

\begin{figure}
	\centering
	\includegraphics[width=0.9\textwidth]{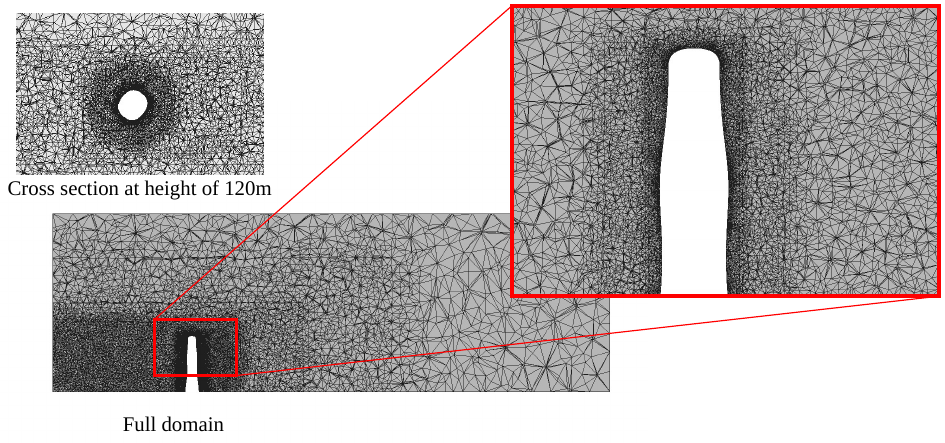}
	\caption{Cross sections of the computational mesh around the building.}
	\label{fig:MeshDetails}
	\end{figure}

\Cref{fig:MeshDetails} depicts specific details of the computational mesh. We note that the domain is meshed with tetrahedral elements with a gradually higher mesh density towards the building. The total number of elements is roughly $7\times 10^5$. The inlet wind is a superposition of the mean profile $\overline{\bmu}$ and the turbulent fluctuations $\bmu^\prime$ described in the previous section; cf. \Cref{fig:WindSnapshots}. For each wind direction $\theta$, the building is rotated inside the domain and the domain is then locally remeshed so that the inflow is at the intended orientation. Body-fitted remeshing is also performed each time the building geometry is updated during optimization.
The parameter values for the fluid and flow domain used in our simulations are tabulated in \Cref{table:fluid_properties}. 

\begin{table}
  \caption{Fluid and flow domain parameters.}
  \centering{}
  \begin{tabular}{lll}
    \toprule
    Parameter               & {Value}  & {Unit}                          \\
    \midrule
    Density                & $1.225$    & kg/m$^3$       \\
        Dynamic viscosity ($\nu$)      & $1.846\times 10^{-5}$ &  Ns/m$^2$ \\
    Reynolds  number ($Re$)      & $2.9\times 10^7$    &                                 \\
        Height of the building ($H$) & $180 $     &  m                           \\
            Time window   & $[50,200]$       &  s                           \\
    \bottomrule
  \end{tabular}
  \label{table:fluid_properties}
\end{table}

The high Reynolds number (cf. \Cref{table:fluid_properties}) makes the flow significantly turbulent. The wind flow around the building is computed by a large eddy simulation (LES) with Kratos Multiphysics. \rev{Kratos Multiphysics uses linear tetrahedral finite elements to discretize both the pressure and the velocity field.
The variational multiscale (VMS) method is then used to stabilize this discretization as described in~\cite{Jordi2016}.
The VMS method relies on two mesh-dependent stabilization parameters; one for the stabilization of the convection part of the PDE and another one for the equal-order velocity and pressure fields. }
The fluid domain is modeled with a fractional step method with a second-order backward differentiation formula (BDF) time integration scheme.
The time step of the BDF scheme is chosen so that the Courant--Friedrichs--Lewy (CFL) number remains less than one in the smallest element near the building domain.
We refer the reader to~\cite{Jordi2016, Andre2015,Codina2001} for further details on our finite element discretization.

The initial condition for the wind, i.e., $\bmu|_{t=0}$, is a zero velocity field throughout the entire of the fluid domain $D\setminus B$.
Starting from this steady state, a total time of $200$ seconds of wind flow is then simulated. This number was chosen so that least 10-15 cycles of vortex shedding are captured. To stabilize the early part of the unsteady simulation, the magnitude of the inflow velocity is rescaled by $t/10$ between $t=0$ and $t=10$. In order to filter transient effects resulting from the zero initial condition initializing the simulation, the first $50$ seconds of the simulation are disregarded, and we only use the flow field data for $t\in[50,200]$.\footnote{\rev{Ten-minute simulations are widely used in computational wind engineering applications \cite{Tosi_2021}. However, we exploit that fact that this simulation time can be reduced through ensemble averaging \cite{MAKARASHVILI2017236,Luo2018,Tosi_2021}.}}

\subsection{Design parameters} \label{sub:design_spce}

\begin{figure}
	\centering
	\includegraphics[trim=0cm 4.6cm 0cm 0cm, clip=true,width=0.8\textwidth]{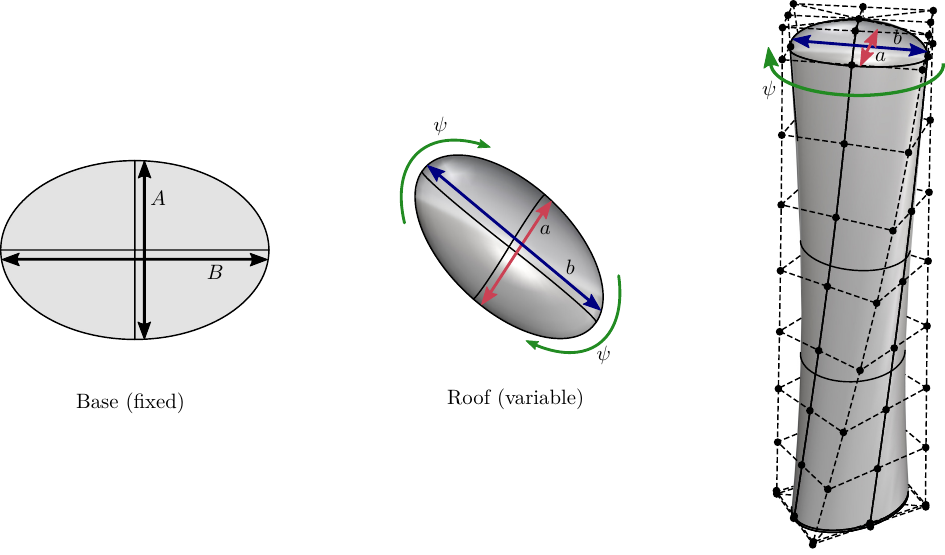}
	\caption{The design parameters $\mathcal{Z} = (a,b,\psi)$ to be optimized. Building as seen from the bottom (left) from the roof (middle), and in 3D (right). The roof area is constant in the numerical example presented; i.e., $\pi a b /4 = \text{const.}$}
	\label{fig:design_spce}
\end{figure}

The parameterization of the building geometry is described in \Cref{sub:details_numerical_model}.
We note that the height of the building can not be altered in a typical design scenario. The base dimensions of the building are fixed due to the fact that the built-up area purchased for construction would be fixed in practice. Hence, in our numerical examples, we keep the base dimensions and the height constant. The cross-sectional area at the roof of the building is also kept constant in order to control the usable area of the building. Major ($b$) and minor diameter ($a$) are hence constrained by the equation $\pi ab/4 = \mathrm{const.}$
As such, the optimizable design space ($\mathcal{Z}$) reduces to two parameters: the minor axis diameter ($a$) and the twist of the building at the roof cross-section ($\psi$); i.e., $\mathcal{Z} = (\psi,a)$. This constrained design space is depicted in \Cref{fig:design_spce}. In other design scenarios, there may be additional design parameters, but we consider only the scenario above in this work.

\subsection{Observables} \label{sub:observables}
There are two main criteria that need to be considered in the design of tall buildings: strength and serviceability criteria.
Strength criteria deal with the strength of the structure and guarantee that the building will not fail under the maximum design load. The serviceability criteria deal with the comfort of the occupants in the building.
For strength criteria, the quantities of interest are based on the total reaction forces and the base moments. Base moment refers to the moment of the forces at the base of the building.
For serviceability criteria, the quantities of interest are the displacement and acceleration of the building.

The quantity of interest that we consider in the current study is the norm of the base moment.
This quantity is of particular interest for the design of the foundation and the structural system of the building.
The mechanical moment created at the center $\bmO$ of the base of the building $B_{\mathcal{Z}}$ by the fluid pressure is defined as
\begin{equation}
  \label{eq:base_moment}
  {M}_{\mathcal{Z}} = \Big\| \int_{B_{\mathcal{Z}}} ({\bmx}-\bmO) \times p({\bmx})\bmn({\bmx}) dS({\bmx}) \Big\| 
  ,
\end{equation}
where $\bmn(\bmx)$ is the unit normal to the building surface.
By minimizing a measure of risk associated to this random variable, a reduction in the total cost of the building can be achieved.

\section{Risk measurement and risk-averse optimization} \label{sec:risk_measures}
\label{sec:chance_constraints}

In this section, we define two risk measures, the expected value and the conditional value-at-risk~\cite{rockafellar2000optimization, rockafellar2015engineering}, in the context of the transient simulations described above.
We then use these risk measures to formulate risk-neutral and risk-averse design optimization problems for tall buildings.

\subsection{A note on averages} \label{sub:averages}

As above, let $\bmu$ and $p$ denote the fluid velocity and pressure, respectively, and $X(t) = \varphi(\bmu(t),p(t))\in \R$ be a system observable at time $t$; cf.~\cref{eq:base_moment}.
We require more than one notion of average to investigate the random observables in this work.
The first notion is the mean with respect to an invariant probability measure $\bbP_t$ at time $t$.
In order to express this average properly, we make the assumption that $\{X(t) \colon t \in (0,\infty)\}$ is a stochastic process where each sample path $X(t) = X(t;\omega)$ is indexed by $\omega \in \Omega$.
The mean of $X$, with respect to $\bbP_t$, at time $t$, is defined
\begin{equation}
			\langle X(t)\rangle
	=
	\int_\Omega X(t;\omega) \dd\bbP_t(\omega)
						\,.
\label{eq:ProbabilityWeightedAverage}
\end{equation}

For \emph{statistically stationary} processes, i.e., stochastic processes whose joint cumulative distribution function (CDF) does not change when translated in time, the integral in~\cref{eq:ProbabilityWeightedAverage} is constant for all time $t$.
It is often assumed that there is a unique measure $\bbP_\star$ such that $\bbP_t = \bbP_\star$ for all $t$ and, furthermore, that for almost every sample path $X(\cdot,\omega)$,~\cref{eq:ProbabilityWeightedAverage} may be rewritten as the (infinite) temporal average\footnote{Note that the lower bound, $T_1$, is arbitrary.}
\begin{equation}
	\langle X \rangle
			=
		\lim_{T\to\infty} \frac{1}{T} \int_{T_1}^{T_1+T} X(t;\omega) \dd t
					\,.
\label{eq:TemporalMean}
\end{equation}
The validity of~\cref{eq:TemporalMean} is an open problem in all but some special situations (cf. \cite{hairer2006ergodicity}).
Nevertheless, the \emph{ergodic hypothesis}~(i.e., \cref{eq:TemporalMean}) remains widely accepted in the fluid dynamics community for many problem types, and we will make use of it.

To compress the subsequent notation, we will denote $\bxi = (u_\ast,\theta,z_0,r)$ and use $\varrho(\bxi)$ to denote the joint probability density function (PDF) of its components.
It is evident that our system observables depend on $\bxi$, $X(t) = X(t,\bxi)$, or, in other words, for every outcome $\bxi$, we will encounter a different stochastic process $\{X(t,\bxi) \colon t \in (0,\infty)\}$.\footnote{In a further abuse of notation, each sample path in the process should be written $X(t,\bxi;\omega)$.}

The presence of $\bxi$ necessitates a second notion of average.
To this end, the mean of $X(t,\bxi)$, with respect to the uncertain parameter vector $\bxi = (u_\ast,\theta,z_0,r)$, is defined
\begin{equation}
	\rmE\big[ X \big]
	=
	\int_0^1 \int_{z_\rmL}^{z_\rmR} \int_0^{2\pi} \int_0^\infty
	X(\cdot,\bxi)
	\varrho(\bxi)
	\dd u_\ast \dd \theta \dd z_0 \dd r
	\,.
\label{eq:ParametricMean}
\end{equation}
Clearly, $\rmE\big[ X \big] (t)$ is a stochastic process in the variable $t$.
Combining the two notions of average,~\cref{eq:TemporalMean,eq:ParametricMean}, we arrive at what we hereafter refer to as the \emph{expected value} of $X$:
\begin{equation}	
	\bbE\big[ X \big]
	=
	\rmE\big[ \langle X \rangle \big]
	\,.
\label{eq:TotalMean}
\end{equation}

It is fundamental for us to estimate the expected value of observables $X(t,\bxi)$.
First, we introduce the (finite) temporal average,
\begin{equation}
	\langle X(\cdot,\bxi) \rangle_{T}
		=
	\frac{1}{T}
	\int_{T_1}^{T_1+T} 
	X(t,\bxi) \dd t
		,
	\qquad T>0
	.
\end{equation}
Second, we introduce the (finite) ensemble average,
\begin{equation}
	\rmE_{S}\big[X\big] = \frac{1}{N}\sum_{i=1}^{N} X(\cdot,\bxi_i)
	,
	\qquad N = |S|>0
	,
\label{eq:TotalMeanMC}
\end{equation}
where $S = \{\bxi_i\}_{i=1}^N$ is a finite sample set of i.i.d. realizations of the random variable $\bxi$.
In this work, the estimator
\begin{equation}	
		\rmE_{S}\big[
	\langle
	X
	\rangle_{T}
	\big]
	\approx
	\bbE[X]
	\,,
\label{eq:SampleAverageLimit_WithTime}
\end{equation}
will be our approximation of choice for estimating~\cref{eq:TotalMean}.

Now that we have defined the expected value, we may immediately define other statistics.
For instance, we define the variance as follows:
\begin{equation}
	\mathrm{Var}[X]
	=
	\bbE\big[
		( X - \bbE[X] )^2
	\big]
	,
\label{eq:Variance}
\end{equation}
and, likewise, the standard deviation $\sigma = \sqrt{\mathrm{Var}[X]}$.
The variance and standard deviation both measure how spread out realizations of the observable $X$ are with respect to time $t$ and the parameters in $\bxi$.
We may use the variance~\cref{eq:Variance} to derive an expression of the variance of the estimator $E_{S}\big[\langle X \rangle_T\big]$.
Indeed, a straightforward computation shows that
\begin{equation}
			\bbE\big[ (E_{S}\big[\langle X \rangle_T\big] - \bbE[X])^2 \big]
	=
	\frac{\mathrm{Var}[\langle X \rangle_T]}{|S|}
	\,.
\label{eq:StatisticalError}
\end{equation}
In general, the numerator $\mathrm{Var}[\langle f \rangle_T]$ will decrease as $T\to\infty$.
However, because of the presence of the random vector $\bxi$, $\mathrm{Var}[\langle f \rangle_T]$ cannot be expected to vanish in the $T\to\infty$ limit \cite{Tosi_2021}.

\begin{figure}
    \centering
    \includegraphics[width=0.9\textwidth]{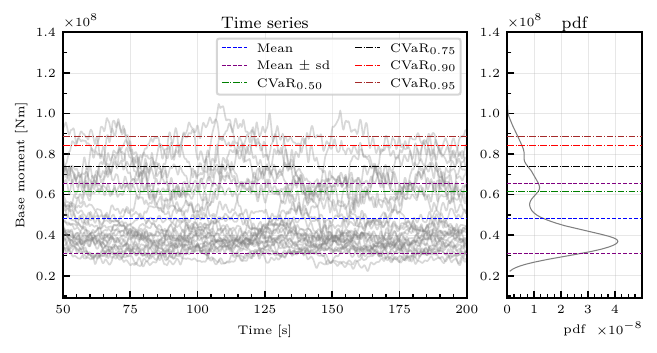}
\caption{Comparison of the expected value, variance, and conditional value-at-risk $\mathrm{CVaR}_{\beta}$ for the base moment $M_{\mathcal{Z}}(t,\bxi)$ when $\mathcal{Z} = \mathcal{Z}_0$ in Numerical study II ~\Cref{sec:results_and_candidate_designs}.}
\label{fig:result_timeseries_single}
\end{figure}

A robust building design should have a low probability of extreme limit states.
Therefore, a robust building design may have a low variance in a random load $X$ simply because a low variance implies a low probability of extreme $X$-values.
Nevertheless, directly controlling the variance/standard deviation is not optimal for our purposes, and we choose to use an alternative measure of risk.
One important reason for seeking alternative risk measures is that $\mathrm{Var}[X]$ penalizes variation both below and above the mean $\bbE[X]$.
Meanwhile, in typical practice, only extreme values on one side of the mean need to be penalized; cf.~\Cref{fig:result_timeseries_single}.
Numerous other drawbacks of optimizing for the variance and standard deviation are outlined in detail in \cite{rockafellar2015engineering}.
As an alternative to $\mathrm{Var}[X]$, we consider the conditional value-at-risk~\cite{rockafellar2000optimization, rockafellar2015engineering}.

\subsection{Conditional value-at-risk} \label{sub:conditional_value_at_risk}

We begin with the definition of the value-at-risk.
Let $F_X(x) := \bbP(X\leq x)$ denote the CDF of a real-valued random variable $X$ defined on a probability space $(\Omega,\scA,\bbP)$.
The value-at-risk ($\mathrm{VaR}$) of $X$, at confidence level $0<\beta<1$, also known as the $\beta$-quantile, is defined
\begin{equation}
	\mathrm{VaR}_\beta(X)
	:=
	\inf\,\{s\in\R \,:\, F_X(s) \geq \beta\}
	\,.
\label{eq:VaR}
\end{equation}
The conditional value-at-risk ($\mathrm{CVaR}$) of $X$, at confidence level $\beta$, is the expected value of $X$ in the largest $(1-\beta)\cdot 100$ percent of possible events.
Indeed, if $X\in L^1(\bbP)$ and $F_X(x)$ is continuous, then $\mathrm{CVaR}$ is precisely the conditional expectation
\begin{equation}
	\mathrm{CVaR}_\beta(X)
	:=
	\bbE[X| X>\mathrm{VaR}_\beta(X)]
	\,.
\end{equation}
From this definition, we find that
\begin{equation}
	\lim_{\beta \to 0} \mathrm{CVaR}_\beta(X) = \bbE[X]
	.
\end{equation}
Alternatively, one may define $\mathrm{CVaR}_\beta(X)$ as the solution of a scalar optimization problem \cite{rockafellar2000optimization}; namely,
\begin{align}
	\mathrm{CVaR}_\beta(X)
	&=
	\frac{1}{1-\beta}
	\int_\beta^1
	\mathrm{VaR}_\alpha(X)
	\dd\alpha
			=
	\inf_{s\in\R}
	\Big\{
		s + \frac{1}{1-\beta}\bbE[(X-s)_+]
	\Big\}
	\,,
\label{eq:CVaR}
\end{align}
where $(x)_+ := \max\{0,x\}$.
We will exploit this latter definition in~\Cref{sub:optimization_problems}.

The $\mathrm{CVaR}$ has many appealing mathematical properties that have been documented in the literature \cite{rockafellar2000optimization,rockafellar2015engineering,kouri2016risk,kouri2018optimization}.
In our setting, we emphasize that $\mathrm{CVaR}_\beta(X)$ is a more useful measure of risk than the mean, variance, or standard deviation because it directly measures the weight of the tail of $X$.
In other words, the $\mathrm{CVaR}$ allows us to measure and optimize for the expected value of the limit states that typically induce failure.

In \Cref{fig:result_timeseries_single}, we compare the expected value, variance, and $\mathrm{CVaR}$ for the base moment $X = M_{\mathcal{Z}}(t,\bxi)$ estimated from $30$ independent samples of $\bxi = (u_\ast,\theta,z_0,r)$ at the initial design state $z = z_0$ in Study II ~\Cref{sec:results_and_candidate_designs}.
From this figure, we see that the initial base moment distribution is multi-modal and highly skewed towards large values.
It is the weight of this tail that we reduce when optimizing for $\mathrm{CVaR}_\beta$.

\subsection{Optimization problems} \label{sub:optimization_problems}

In this paper, we focus on three optimization problems written as follows:
\begin{gather}
\tag{Prob.~1}
\label{eq:prob1}
	\min_{\mathcal{Z}\in C}
	\Big\{
		J(\mathcal{Z})
		:=
		\bbE[ M_{\mathcal{Z}} ]
	\Big\}
	,
	\\[3pt]
\tag{Prob.~2}
\label{eq:prob2}
	\min_{\mathcal{Z}\in C}
	\Big\{
		J(\mathcal{Z};\beta)
		:=
		\mathrm{CVaR}_\beta [M_{\mathcal{Z}}]
	\Big\}
	, 
	\\[3pt]
\tag{Prob.~3}
\label{eq:prob3}
	\min_{\mathcal{Z}\in C}
	\Big\{
		J(\mathcal{Z})
		:=
		\langle M_{\mathcal{Z}} \rangle
	\Big\}
	\quad
	\text{subject to }
	\bxi
	=
	\bxi_{\mathrm{PWD}}
	.
\end{gather}
As before, $\mathcal{Z}$ denotes the design variable, and $C$ denotes the design space.
From now on, we set $\beta = 0.90$. $\bxi_{\mathrm{PWD}}$ corresponds to the predominant wind direction (PWD), $\theta = \theta_{\mathrm{PWD}}$ and with all remaining mean wind field parameters at their mean values. 
The optimum of~\ref{eq:prob1} is a \emph{risk-neutral} design that has the lowest expected value of the base moment.
The optimum of~\ref{eq:prob2} is a \emph{risk-averse} design that has the lowest $10\%$-tail expectation of the base moment.
The optimum of~\ref{eq:prob3} is a design that has the lowest time average of base moment for $\bxi=\bxi_{\mathrm{PWD}}$.
This optimization problem does not take the uncertainty into consideration.
Therefore, it is essentially a deterministic optimization problem.
In special cases, the optimal designs for all three problems,~\ref{eq:prob1},~\ref{eq:prob2}, and~\ref{eq:prob3} can be close to each other.
However, for complicated geometries and environments, the resulting designs may be significantly different; cf. \cite[Section~6.3]{beiser2020adaptive}.

\section{Adaptive stochastic optimization algorithm} \label{sec:optimization_algorithm}

In this section, we describe the iterative, gradient-based, adaptive stochastic optimization algorithm we have used to solve~\ref{eq:prob1} and~\ref{eq:prob2}.
Our algorithm reduces the cost of optimization by adjusting the number of simulations $N = N_k$ in each gradient estimate based on the accuracy of the current design iterate $\mathcal{Z}_k$ similar to \cite{byrd2012sample,bollapragada2018adaptive,xie2020constrained,beiser2020adaptive}.

\subsection{Monte Carlo approximation of the objective functions} \label{sub:approximation_of_objective_functions}

We have already seen how to approximate $J(\mathcal{Z})$ in~\Cref{sec:chance_constraints}.
Indeed, invoking~\cref{eq:SampleAverageLimit_WithTime}, we have that $J(\mathcal{Z}_k) \approx J_{S_k}(\mathcal{Z}_k)$ where $S_k = \{\bxi_i\}_{i=1}^{N_k}$,
\begin{equation}
		J_{S_k}(\mathcal{Z}_k)
	=
							\frac{1}{N_k}
	\sum_{i=1}^{N_k}
	J_i(\mathcal{Z}_k)
	,
\label{eq:ObjectiveFunctionApproximation}
\end{equation}
and
\begin{equation}
	J_i(\mathcal{Z}) =
	\frac{1}{T}
	\int_{T_1}^{T_1+T} 
	M_{\mathcal{Z}}(t,\bxi_i)
	\dd t
		\,.
\label{eq:ObjectiveFunctionSample}
\end{equation}
The gradient of each sample of the objective function, $J_i(\mathcal{Z})$, can be computed via finite differences in the design variable $z \in C \subset \R^d$ at a cost of $d+1$ independent numerical wind tunnel simulations.\footnote{
Due to the chaotic nature of the flow, an accurate gradient for an individual sample requires a long time interval \cite{lea2000sensitivity}, the exact length of which is problem-dependent.
In addition, special care must be taken to select an appropriate finite difference increment in each component of design space.
}
The empirical mean of these sample gradients is an approximation of the gradient of $J(\mathcal{Z})$, namely,
\begin{equation}
	\nabla J(\mathcal{Z}_k) \approx
	\nabla J_{S_k}(\mathcal{Z}_k)
	=
							\frac{1}{N_k}
	\sum_{i=1}^{N_k}
	\nabla J_i(\mathcal{Z}_k)
	,
\label{eq:ObjectiveFunctionGradientApproximation}
\end{equation}
where, $\nabla$ denotes the gradient. The gradients are approximated via finite differences with respect to the design variable $\mathcal{Z}$.

The modifications involved in generalizing \cref{eq:ObjectiveFunctionApproximation,eq:ObjectiveFunctionGradientApproximation} to the objective function $J(\mathcal{Z};\beta)$ require that we invoke~\cref{eq:CVaR} to rewrite
\begin{equation}
		J(\mathcal{Z};\beta)
	=
		\min_{s\in\R}
	\Big\{
		s + \frac{1}{1-\beta}\bbE[(M_{\mathcal{Z}}-s)_+]
	\Big\}
	.
\end{equation}
We now define the following generalization of~\cref{eq:ObjectiveFunctionSample}:
\begin{equation}
	J_i(\mathcal{Z},s)
	=
	\frac{1}{T}
	\int_{T_1}^{T_1+T} 
		\big(M_{\mathcal{Z}}(t,\bxi_i) - s\big)_+
	\dd t
		\,.
\label{eq:ObjectiveFunctionSampleCVaR}
\end{equation}
Then, after defining $s^\star = s^\star(\mathcal{Z}_k,S_k,\beta)$ via the one-dimensional minimization problem
\begin{equation}
	s^\star
	=
	\argmin_{s\in\R}
	\Big\{
		s + \frac{1}{1-\beta}\frac{1}{N_k}\sum_{i=1}^{N_k} J_i(\mathcal{Z}_k,s)
	\Big\}
	,
\end{equation}
we approximate $J(\mathcal{Z}_k;\beta)\approx J_{S_k}(\mathcal{Z}_k;\beta)$, as well as its gradient $\nabla J(\mathcal{Z}_k;\beta)\approx \nabla J_{S_k}(\mathcal{Z}_k;\beta)$, as follows:
\begin{equation}
\label{eq:CVaRSAA}
	J_{S_k}(\mathcal{Z}_k;\beta)
	=
	\frac{1}{N_k}
	\sum_{i=1}^{N_k}
	J_i(\mathcal{Z}_k,s^\star)
	,
	\qquad
	\nabla J_{S_k}(\mathcal{Z}_k;\beta)
	=
	\frac{1}{N_k}
	\sum_{i=1}^{N_k}
	\nabla J_i(\mathcal{Z}_k,s^\star)
	.
\end{equation}
From now on, for notational simplicity, let $J(\mathcal{Z})$ denote both $J(\mathcal{Z})$ and $J(\mathcal{Z};\beta)$ in \ref{eq:prob1} and \ref{eq:prob2}.
Likewise, we will write $J_{S_k}(\mathcal{Z})$ for both $J_{S_k}(\mathcal{Z})$ and $J_{S_k}(\mathcal{Z};\beta)$.

Once the gradients above have been computed at $\mathcal{Z}_{k}$, we apply the stochastic gradient descent method \cite{wright2022optimization} to define the next iterate,
\begin{equation}
	\mathcal{Z}_{k+1}
	=
	\mathcal{Z}_k - \alpha \nabla J_{S_k}(\mathcal{Z}_k)
	,
	\qquad
	\alpha>0
	\,.
\label{eq:SGD}
\end{equation}
For greater robustness and efficiency, we adaptively select each batch size $N_k = |S_k|$ based on an a posteriori estimate of the statistical error described in the next subsection \cite{byrd2012sample,bollapragada2018adaptive,bollapragada2019adaptive,beiser2020adaptive,xie2020constrained}.

\begin{remark}
	Owing to the presence of the non-smooth operator $(\,\cdot\,)_{+}$ in~\cref{eq:ObjectiveFunctionSampleCVaR}, the functions $J_i(\mathcal{Z},s)$ are not continuously differentiable with respect to $z$ or $s$.
	Even though the gradient $\nabla J_i(\mathcal{Z},s)$ can be computed uniquely at almost every design point $x$, the non-differentiability can present issues if a naive gradient descent algorithm is used \cite{rockafellar2000optimization,kouri2016risk}.
	In our simulations, characteristic issues with gradient descent did not appear, likely because of the step size we used was small enough that the optimization errors remained lower than other simulation errors.
	\end{remark}

\subsection{Adaptive sampling} \label{sub:adaptive_sampling}

We may note from~\cref{eq:StatisticalError} that the variance in the gradient estimator $\nabla J_{S_k}(\mathcal{Z}_k)$ is inversely proportional to $N_k = |S_k|$.
Therefore, a large batch size $N_k$ at each stochastic gradient descent iteration~\cref{eq:SGD} will lead to a high probability of decreasing the objective function, i.e., $J(\mathcal{Z}_{k+1}) < J(\mathcal{Z}_k)$.
This, in turn, will reduce the total number of iterations necessary to optimize the design.
On the other hand, using a large batch size at every iteration is unnecessarily costly with respect to the number of samples.
It turns out that linear convergence can be achieved by starting with a small batch size that grows as $\|\nabla J(\mathcal{Z}_k)\| \to 0$.

In this work, we use an adaptive sampling strategy based on the ``norm test'' introduced in \cite{byrd2012sample,bollapragada2018adaptive} in order to tune the batch size.
The strategy consists of the following steps.
At the outset, a relatively small batch of samples $S_0$ is chosen and, before each subsequent iteration $k+1\geq 0$, an assessment is made whether the computed gradient is likely to reduce the objective function.
If it is judged that the accuracy of the gradient is sufficient, the next batch will have the same size, i.e., $|S_{k+1}| = |S_k|$; otherwise, a larger batch size will be chosen at the next iteration, i.e., $|S_{k+1}| > |S_k|$.

The norm test delivers a posteriori control of the variance of the sample gradient $\nabla J_{S_k}(\mathcal{Z}_k)$.
It is built around the observation that $\nabla J_{S_k}(\mathcal{Z}_k)$ is a descent direction $\mathcal{Z}_k$, for sufficiently smooth $J$, if
\begin{equation}
	\|\nabla J_{S_k}(\mathcal{Z}_k) - \nabla J(\mathcal{Z}_k)\|^2
	\leq
	\vartheta^2
	\|\nabla J_{S_k}(\mathcal{Z}_k)\|^2
	,
	\quad
	\text{for some }
	\vartheta\in (0,1)
	.
\end{equation}
Computing the left-hand size exactly is infeasible, but if we replace the expression with its expectation, i.e.,
\begin{equation}
	\bbE\big[\|\nabla J_{S_k}(\mathcal{Z}_k) - \nabla J(\mathcal{Z}_k)\|^2\big]
	=
	\mathrm{Var}(\nabla J_{S_k}(\mathcal{Z}_k))
	=
	\frac{\mathrm{Var}(\nabla J_i(\mathcal{Z}_k))}{|S_k|}
	,
\end{equation}
then it may be accurately estimated.
Indeed, the true variance of the gradient samples, $\mathrm{Var}(\nabla J_i(\mathcal{Z}_k))$, can be approximated by the sample variance
\begin{equation}
	\mathrm{Var}_{S_k}
	(\nabla J_i(\mathcal{Z}_k))
	=
	\frac{1}{|S_k|-1} 
	\sum_{i=1}^{N_k}
	\|\nabla J_i(\mathcal{Z}_k) -\nabla J_{S_k}(\mathcal{Z}_k)\|^2
	.
\end{equation}
Using this expression, we arrive at the norm test:
\begin{equation}
\label{eq:NormTest}
	\frac{\mathrm{Var}_{S_k}(\nabla J_i(\mathcal{Z}_k))}{|S_k|}
	\leq
	\vartheta^2\|\nabla J_{S_k}(\mathcal{Z}_k)\|^2
	\,.
\end{equation}
It has been shown that an idealized norm test gives optimal convergence rates for convex objective function and is robust enough to efficiently deal with many non-convex problems \cite{byrd2012sample,bollapragada2018adaptive}.

At an iteration $k$ where~\cref{eq:NormTest} is violated, the subsequent batch $S_{k+1}$ is prescribed to have a sample size satisfying
\begin{equation}
\label{eq:NewSampleSizeFormula}
	|S_{k+1}|
	=
	\mathrm{Ceiling}\left[
	\frac{\mathrm{Var}_{S_k}(\nabla J_i(\mathcal{Z}_k))}{\vartheta^2\|\nabla J_{S_k}(\mathcal{Z}_k)\|^2}
	\right]
	,
\end{equation}
where $\mathrm{Ceiling}[\cdot]$ returns the smallest integer greater than or equal to its argument.
On the other hand, if~\cref{eq:NormTest} is satisfied, the next batch size remains unchanged.
The entire procedure is summarized in \Cref{alg:AdaptiveSampling}.
\begin{algorithm2e}
\DontPrintSemicolon
	\KwIn{initial design $\mathcal{Z}_0$, initial sample set $S_0$, step size $\alpha > 0$, constant $0 < \vartheta < 1$.}
		\smallskip
	set $k \gets 0$\\
	\Repeat{convergence test is satisfied}{
																						
						compute new design iterate: $\mathcal{Z}_{k+1} = \mathcal{Z}_k - \alpha \nabla J_{S_k} (\mathcal{Z}_k)$\\
		\eIf{condition~\cref{eq:NormTest} is satisfied}{
			compute a new sample set $S_{k+1}$ satisfying $|S_{k+1}| = |S_{k}|$
		}{
			compute a new sample set $S_{k+1}$ satisfying~\cref{eq:NewSampleSizeFormula}
		}
		set $k \gets k+1$\\
	}
\caption{\label{alg:AdaptiveSampling}Stochastic gradient descent with adaptive sampling.}
\end{algorithm2e}

The general benefits of this strategy are three-fold: (1) the small initial sample size allows for fast progress towards the optimal design when one begins with a poor initial guess and the optimization error dominates all other errors; (2) the progressively growing batch sizes allow for a logical growth of samples which keeps the sampling error in line with optimization error; and (3) there is little chance of expensive ``over-sampling'' as the algorithm adaptively chooses the correct sample size for the given problem.
\rev{
The potential benefits of adaptive sampling are even more significant with risk-averse optimization problems such as~\ref{eq:prob2}.
Indeed, when using~\cref{eq:ObjectiveFunctionSampleCVaR} to compute the $\mathrm{CVaR}$-based objective functions, the $(\cdot)_+$ operator zeroes out many of the samples.
This, in turn, increases the problem complexity as it demands more samples to be drawn in order to produce an accurate gradient estimate.
A brief study on the evolution of the sample size as a function of $\beta$ can be found in~\cite{beiser2020adaptive}. 
}

\subsection{Stochastic optimization workflow}
\label{aerodynamic_optimization_workflow}

\begin{figure}
		\centering
		\includegraphics[clip=true, trim=0.75cm 0 0 0, width=\textwidth]{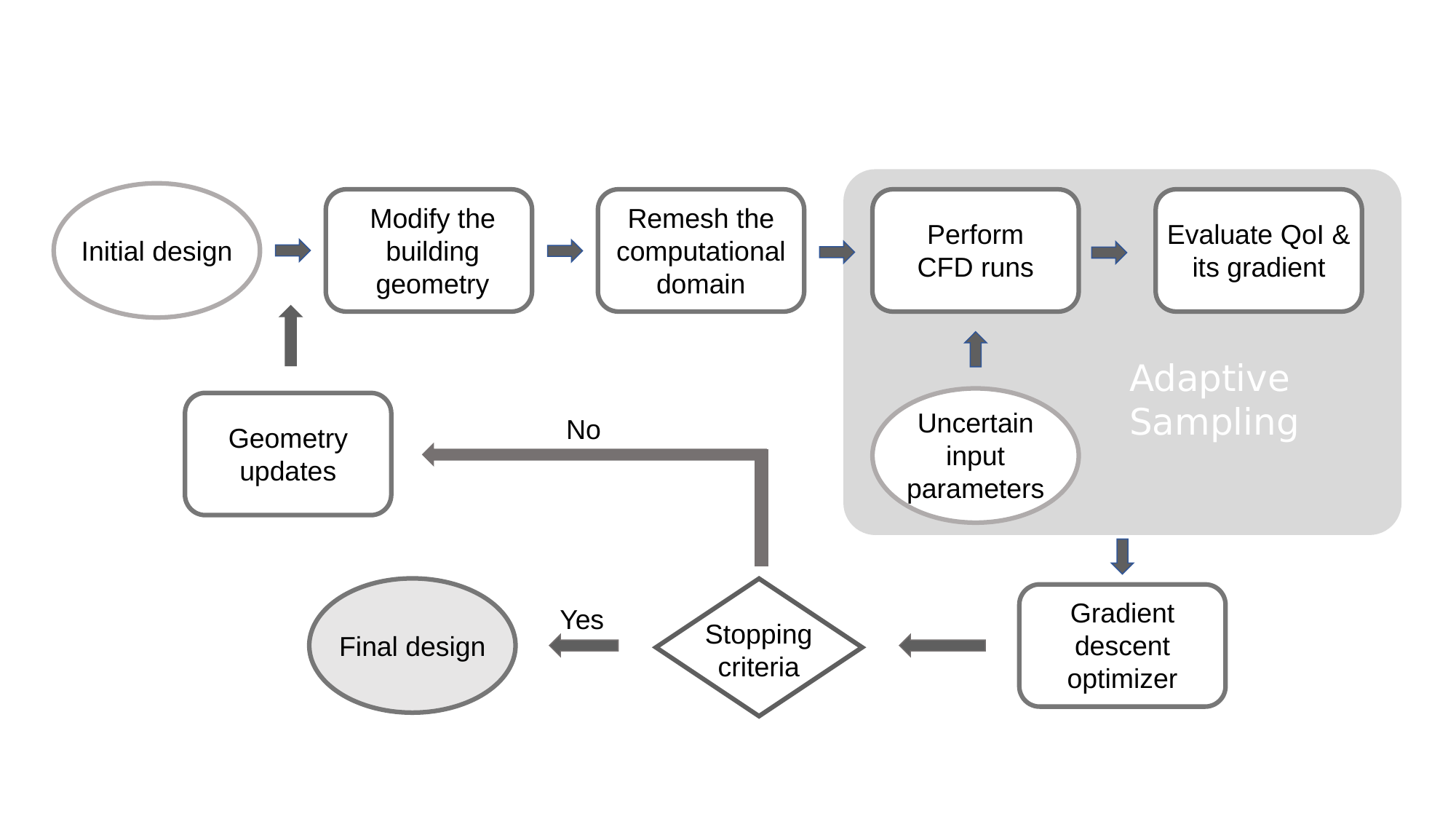}
			\caption{Flow chart depicting the stochastic optimization workflow.}
	\label{fig:aerodynamic_ouu}
\end{figure}

The adopted stochastic optimization workflow is detailed in this section.
The optimization tries to minimize the observables of the selected QoI by looking for the best design parameters while considering the uncertainty from the incoming wind also considering the constraints.  \\
A diagram of the optimization workflow is presented in \Cref{fig:aerodynamic_ouu}.
The workflow begins by defining the objective function $J$, design parameters $\mathcal{Z}$, and uncertain problem variables $\bxi$. The objective function $J$ is based on the base moment $M_{\mathcal{Z}}$ and is defined to be either the expected value $J(\mathcal{Z}) = \bbE[M_{\mathcal{Z}}]$ (cf. \ref{eq:prob1}) or the conditional value-at-risk at confidence level $0.90$, $J(\mathcal{Z}) = \mathrm{CVaR}_{0.90}(M_{\mathcal{Z}})$ (cf. \ref{eq:prob2}). The objective function depends on the geometry of the building (cf. \Cref{fig:tower_setting}), which in turn depends on the design parameters which are denoted by the parameter vector $\mathcal{Z}$.

Each time the CAD building geometry or incidence angle $\theta$ is changed, the background CFD mesh is remeshed (cf. \Cref{fig:MeshDetails}) to capture the new geometry and create the new body-fitted mesh.
Samples of the objective function, $J_i(\mathcal{Z}_k)$, are obtained by simulating the wind flow around the building in Kratos multiphysics \cite{dadvand2010object} (cf. \Cref{fig:WindSnapshots}).
These 3D CFD simulations are expensive so an adaptive sampling strategy (i.e., \Cref{alg:AdaptiveSampling}) is adopted to reduce the number of samples as much as possible.

Finite differences are used to estimate the gradients $\nabla J_i(\mathcal{Z}_k)$.
Since the number of design parameters is low, the additional work required to estimate these gradients is not prohibitive.
Once $\nabla J_i(\mathcal{Z}_k)$ is computed, it is used to update the design via the stochastic gradient descent update rule~\cref{eq:SGD}. The optimization algorithm requires multiple iterations until it converges to the final design. A relative tolerance of 0.01 is chosen to assess convergence of the algorithm.

\rev{
\begin{remark}
	We used finite differences to estimate the gradients of the time-average quanity $J_i(\mathcal{Z}_k) = \frac{1}{T} \int_{T_1}^{T_1+T} (M_{\mathcal{Z}}(t,\bxi_i) - s)_+ \dd t$ in part because the associated adjoint problem is mathematically unstable at high Reynolds numbers~\cite{lea2000sensitivity,wang2013drag}.
	A more conventional workflow would involve an adjoint-based method or algorithmic differentiation to estimate the gradients $\nabla J_i(\mathcal{Z}_k)$.
	Unfortunately, the development of stable and efficient sensitivity methods for turbulent flows remains an open research topic \cite{blonigan2018least,chandramoorthy2019feasibility,chandramoorthy2021probability}.
\end{remark}
}

\section{Results and candidate designs} The optimization algorithm outlined in \Cref{sec:optimization_algorithm} is applied in two numerical studies in this section. Study I uses only one design parameter, $\mathcal{Z} = \psi$, while Study II involves two design parameters, $\mathcal{Z} = (\psi,a)$, under the area constraint $\pi ab/4 = c$, for some fixed constant $c>0$.
The many independent CFD simulations were scheduled using the task scheduler Compass \cite{COMPSsCode2015,Lordan2013,Tejedor2017}. The computations were run on the \href{https://docs.it4i.cz/karolina/introduction/}{Karolina} supercomputer of the \href{https://www.it4i.cz/en}{IT4Innovations} cluster located in Ostrava, Czech Republic.
The entire optimization process is expensive. To give a scale of the cost, the $\mathrm{CVaR}$ optimization in~\Cref{sub:Study2} required more than $4.36\times 10^5$ CPU hours.

\label{sec:results_and_candidate_designs}
\subsection{Numerical study I}

\begin{figure}
    \centering
    \includegraphics[width=0.9\textwidth]{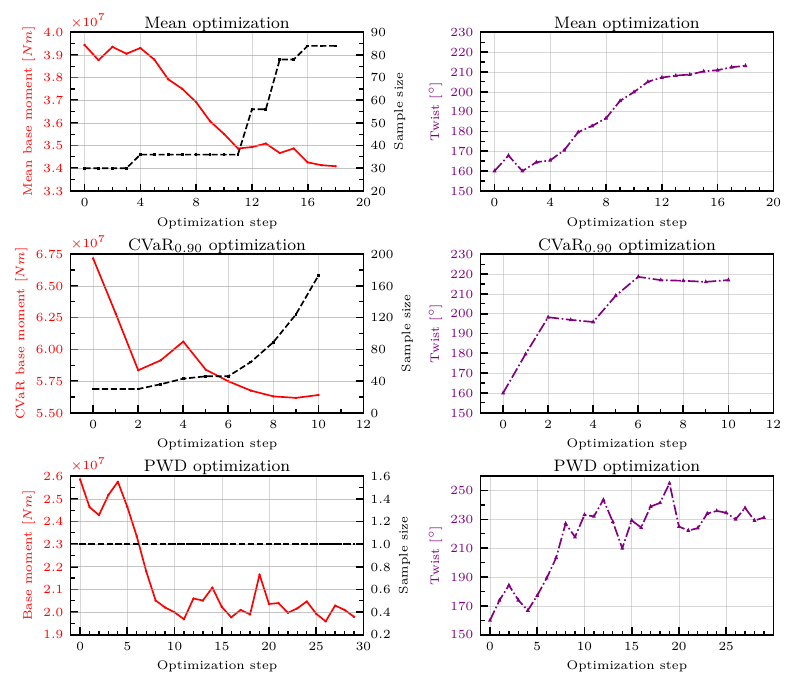}
\caption{Numerical study I: optimization records for the risk-neutral (expected value, \ref{eq:prob1}), risk-averse ($\mathrm{CVaR}$, \ref{eq:prob2}), and PWD problems (\ref{eq:prob3}).}
\label{fig:result_all_one}
\end{figure}

To asses our implementation, we begin with a single parameter optimization study. The details of the initial building geometry are given in \Cref{table:Numerical_study_1}. The only design parameter is the twist of the building $\mathcal{Z} = \psi$; i.e., the major and minor diameters, $a$ and $b$, are kept fixed.
Under this condition, we solve \ref{eq:prob1}, \ref{eq:prob2}, and \ref{eq:prob3} and present the results in \Cref{fig:result_all_one}. In \ref{eq:prob3} we simulate only for the deterministic scenario $\bxi = \bxi_{\mathrm{PWD}}$, which corresponds to the predominant wind direction $\theta = \theta_{\mathrm{PWD}}$ with the remaining wind field parameters fixed at their mean values, $u_\ast = \bbE[u_\ast]$ and $z_0 = \bbE[z_0]$, and for a fixed random seed $\rseed = \rseed_{\mathrm{PWD}}$; i.e.,
\begin{equation}
    \bxi_{\mathrm{PWD}}
    =
    (\bbE[u_\ast],\theta_{\mathrm{PWD}},\bbE[z_0],\rseed_{\mathrm{PWD}})
    .
\end{equation}
In our case (cf. \Cref{table:uncertainties}), $\theta_{\mathrm{PWD}} = 260^{\circ}$, $\bbE[z_0] = 0.05$, and $\bbE[u_\ast] = 10 \mathrm{m}/\mathrm{s}$.
The random seed $\rseed_{\mathrm{PWD}}$ was chosen arbitrarily.

\begin{table}
    \caption{Numerical studies I and II: Initial building design.}
    \centering{}
                                                                        \begin{tabular}{lcc}
      \toprule
                                        & Study I      & Study II     \\
      \midrule
      Angle of twist ($\psi_0$)         & 160$ ^\circ$ & 295$ ^\circ$ \\
      Major diameter at top ($a_0$)   & 35m          & 30m         \\ 
      Minor diameter at top ($b_0$)   & 20m          & 30m         \\
      \bottomrule
    \end{tabular}
    \label{table:Numerical_study_1}
\end{table}

\begin{figure}
    \centering
    \includegraphics[width=0.99\textwidth]{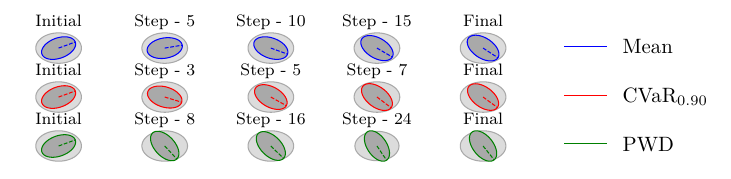}
\caption{Numerical study I: progress of optimization for the risk-neutral, risk-averse, and PWD problems viewed from the top cross section.}
\label{fig:result_all_ellipse_one}
\end{figure}

\begin{figure}
	\centering
	\includegraphics[trim=0cm 5.5cm 0cm 0cm, clip=true,width=1.0\textwidth]{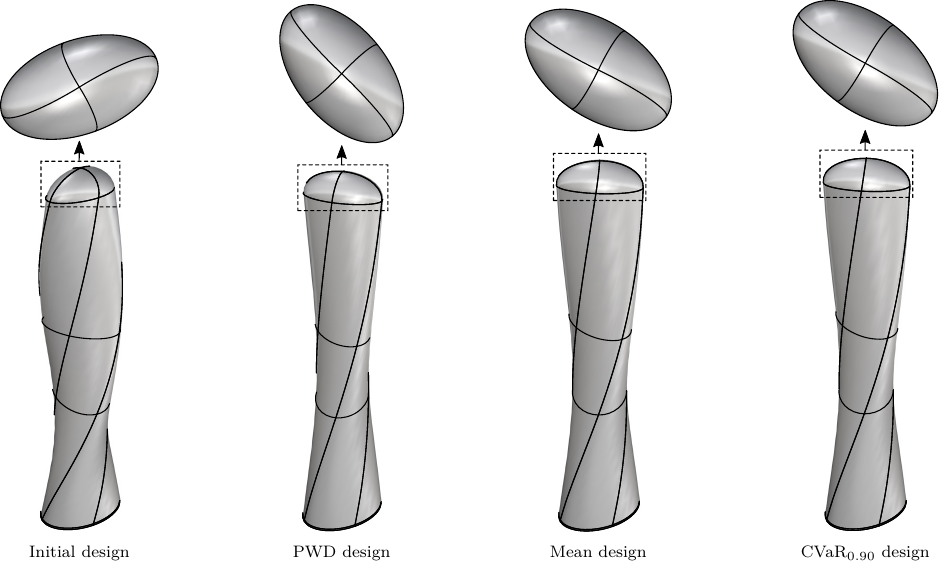}
    \caption{Numerical study I: initial and final building designs.}
    \label{fig:result_geometry_one}
\end{figure}

\begin{figure}
	\centering
	\includegraphics[trim=0cm 4.5cm 0cm 0cm, clip=true,height=0.275\textheight]{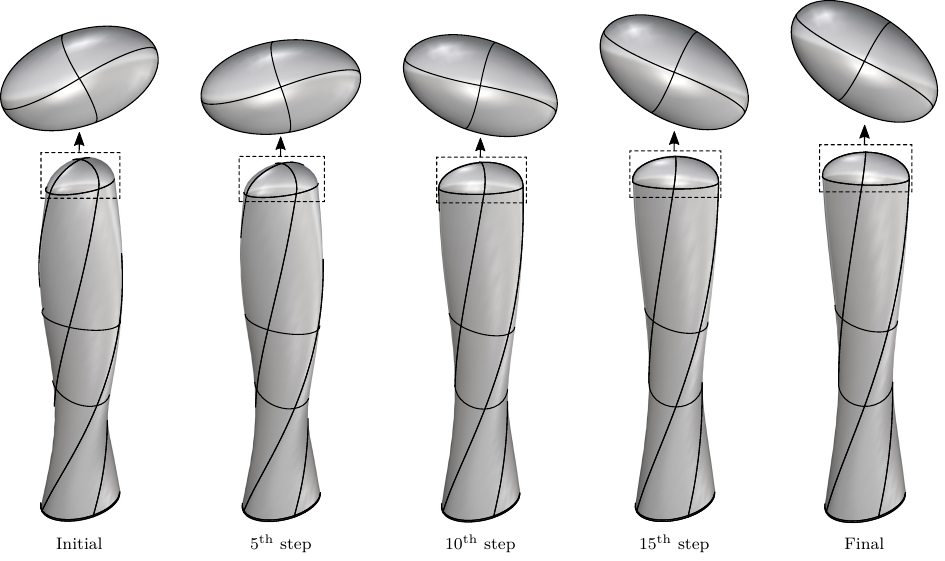}
    \caption{Numerical study I: shape change for risk-neutral base moment optimization \cref{eq:prob1}.}
    \label{fig:result_shape_StudyI_Mean}
\end{figure}

\begin{figure}
	\centering
	\includegraphics[trim=0cm 4.5cm 0cm 0cm, clip=true,height=0.275\textheight]{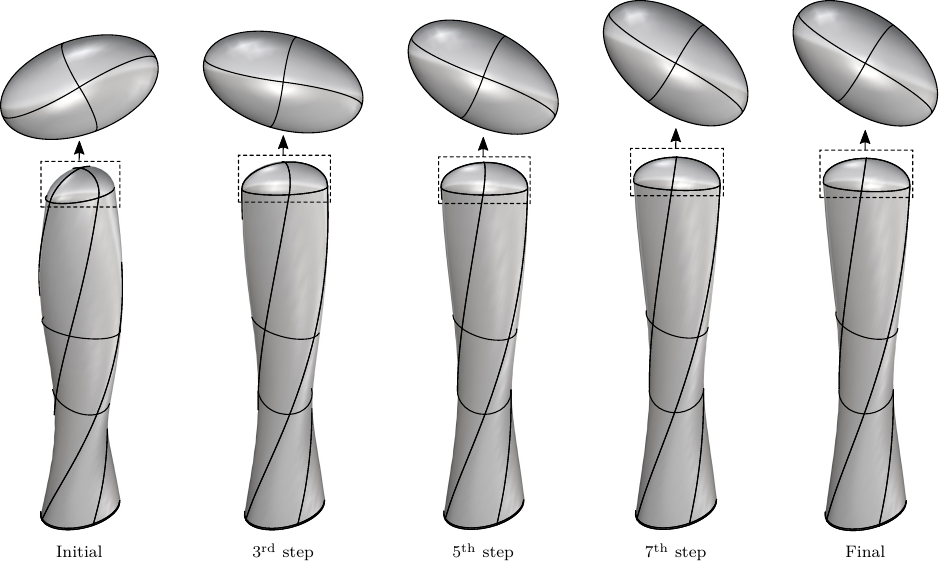}
    \caption{Numerical study I: shape change for risk-averse base moment optimization \cref{eq:prob2}.}
    \label{fig:result_shape_StudyI_CVaR}
\end{figure}

\begin{figure}
    \centering
    \includegraphics[trim=0cm 4.5cm 0cm 0cm, clip=true,height=0.275\textheight]{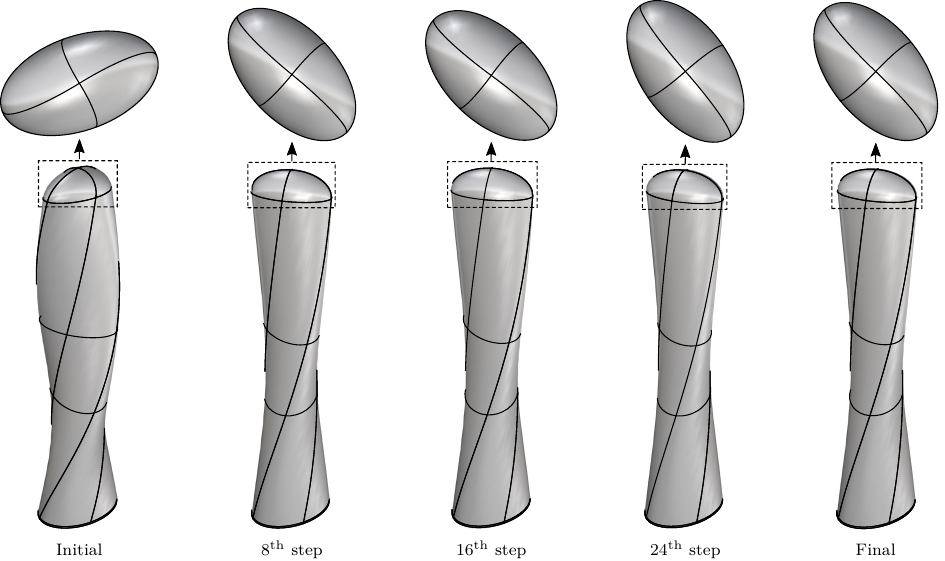}
    \caption{Numerical study I: shape change for PWD optimization \cref{eq:prob3}.}
    \label{fig:result_shape_StudyI_Det}
\end{figure}

 As seen in \Cref{fig:result_all_ellipse_one} and \Cref{table:Results_study_1}, the PWD design differs significantly from the mean and $\mathrm{CVaR}$ designs. The PWD optimization does not take the effects of the uncertain wind (direction, magnitude and fluctuations) into account and, hence, misses important information about the physical environment. \ref{eq:prob1} and \ref{eq:prob2} do consider the variability of the input parameters and, therefore, return more robust designs.
The final building designs are collected together for further comparison in~\Cref{fig:result_geometry_one}.The optimization records of  PWD, risk-neutral,and risk-averse optimization are shown in \Cref{fig:result_shape_StudyI_Det}, \Cref{fig:result_shape_StudyI_Mean}, and \Cref{fig:result_shape_StudyI_CVaR} respectively. \Cref{fig:result_all_ellipse_one} shows the optimization record of all three problems as viewed from top.

\begin{table}
    \caption{Numerical study I: Optimized building designs.}
    \centering{}
    \begin{tabular}{lllll}
      \toprule
      Optimization type             & {$J_{\mathrm{Final}}$} & {$1-\frac{J_\mathrm{Final}}{J_\mathrm{Initial}}$}  & {Twist ($\psi$)}  & $J_{\mathrm{Final},\,\bxi = \bxi_{\mathrm{PWD}}}$    \\
      \midrule
      Risk-neutral \cref{eq:prob1}                &  $34.09\times 10^6$Nm & 13.6\% & 214.08$^\circ$ &  $37.36\times 10^6$Nm \\
      Risk-averse \cref{eq:prob2}       &  $56.40\times 10^6$Nm & 16.0\% & 216.97$^\circ$ &  $58.49\times 10^6$Nm\\
      PWD \cref{eq:prob3}  &  $19.79\times 10^6$Nm & 23.4\% & 231.24$^\circ$ &$19.79\times 10^6$Nm \\
      \bottomrule
    \end{tabular}
    \label{table:Results_study_1}
\end{table}

\begin{figure}
    \centering
    \includegraphics[width=0.9\textwidth]{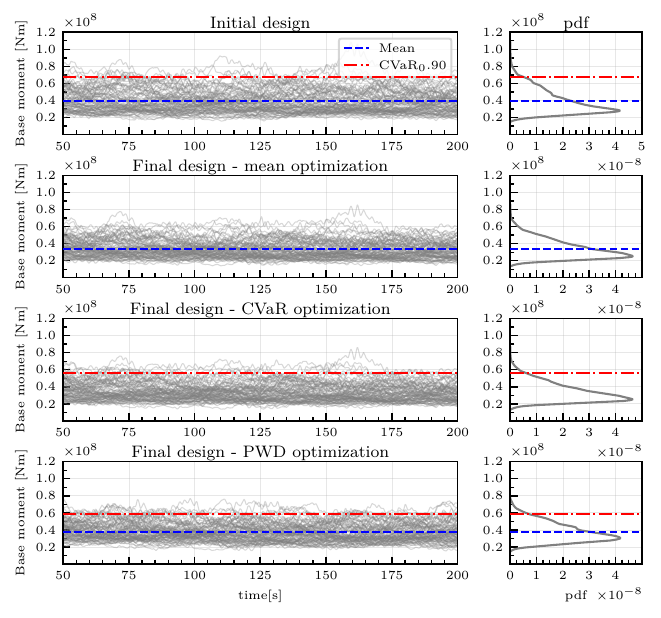}
\caption{Numerical study I: time series and PDFs of the base moment for \ref{eq:prob1},  \ref{eq:prob2}, and  \ref{eq:prob3}.}
\label{fig:result_all_time_series_one}
\end{figure}

Both \ref{eq:prob1} and \ref{eq:prob2} return essentially the same building design.
This is likely due to the extremely low-dimensional nature of the design space.
On the other hand, the final PWD design is far from optimal. The objective functions are compared for the respective PWD values in the last column of \ref{table:Results_study_1}. $\mathcal{Z}_{\mathrm{Final~PWD}}$ represents the final design of PWD optimization. 
The time series and PDFs of the base moments for each optimization problem are shown in \Cref{fig:result_all_time_series_one}. Both of the time series for mean and $\mathrm{CVaR}$ solutions have shifted towards a lower base moment distribution. The performance of the final PWD design considering the uncertainties of the wind is shown. However, the distribution of the PWD design has not improved greatly, when the uncertainty in wind is taken into account, this emphasizes the need for optimization under uncertainties. 
We now turn to a two-parameter optimization problem where the final mean and $\mathrm{CVaR}$ designs significantly differ.

\subsection{Numerical study II}
\label{sub:Study2}

We now consider stochastic optimization of a twisted and tapered building with a fixed roof surface area $c>0$; i.e., $\mathcal{Z} = (\psi,a,b)$ with $\pi ab/4 = c$.
In this study, we consider only risk-neutral base moment optimization \cref{eq:prob1} and the risk-averse base moment optimization \cref{eq:prob2}. The stochastic gradient descent with adaptive sampling (\cref{alg:AdaptiveSampling}) converges after $\sim25$ iterations.

\begin{table}
    \caption{Numerical study II: optimized building designs.}
    \centering{}
    \begin{tabular}{lllll}
      \toprule
      Optimization type     & $J_\mathrm{Final}$  & {$1-\frac{J_\mathrm{Final}}{J_\mathrm{Initial}}$}  & {Twist ($\psi$)}  & {Minor axis length ($a$)}  \\
      \midrule
      Risk-neutral \cref{eq:prob1}        &  $46.36\times 10^6$Nm & $5.811\%$ & $298.019^\circ$ & $24.383$m \\
      Risk-averse \cref{eq:prob2}       &  $69.55\times 10^6$Nm & $17.084\%$ & $307.675^\circ$ & $21.904$m \\
      \bottomrule
    \end{tabular}
    \label{table:Results_study_2}

\end{table}

\begin{figure}
	\centering
	\includegraphics[trim=0cm 4.5cm 0cm 0cm, clip=true,height=0.275\textheight]{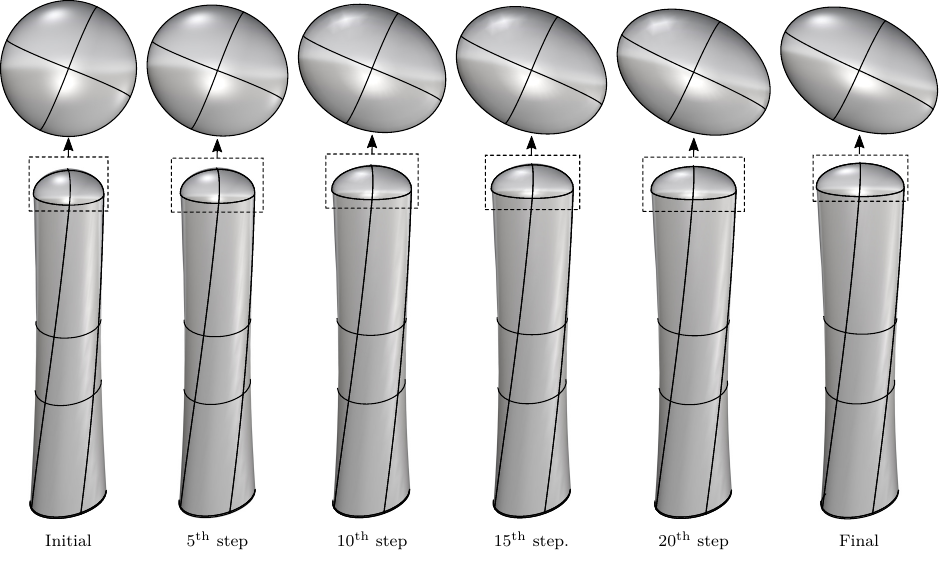}
    \caption{Numerical study II: shape change for risk-neutral base moment optimization \cref{eq:prob1}.}
    \label{fig:result_shape_Mean}
\end{figure}

To avoid significant overlap with Numerical study I, we highlight only the primary features of the risk-neutral candidate design.
It can be seen from \Cref{table:Results_study_2} that the objective function improves by $\sim 6\%$ for this optimization problem.
Although the objective function does not improve greatly, the sequence of designs in \Cref{fig:result_shape_Mean} still clearly illustrates the most important shape changes.
Indeed, the top cross-section becomes more tapered and the twist aligns close to the PWD to reduce the wind effects on the building.
Interestingly, the twisting is not as prominent as the tapering in the final building design.

\begin{figure}
		\centering
		\includegraphics[width=0.8\textwidth]{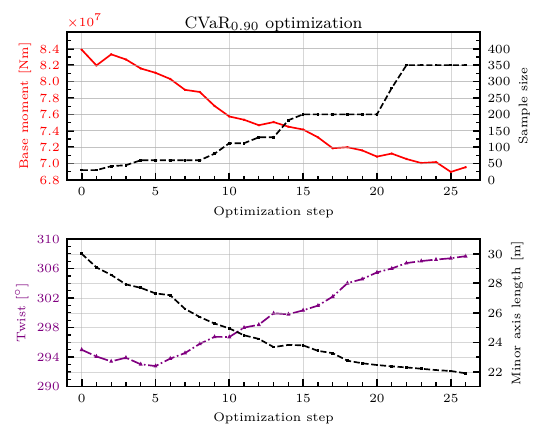}
	\caption{Numerical study II: optimization record for $\mathrm{CVaR}$ problem \cref{eq:prob2}.}
	\label{fig:result_cvar}
\end{figure}

\begin{figure}
	\centering
	\includegraphics[trim=0cm 6.1cm 0cm 0cm, clip=true,width=0.9\textwidth]{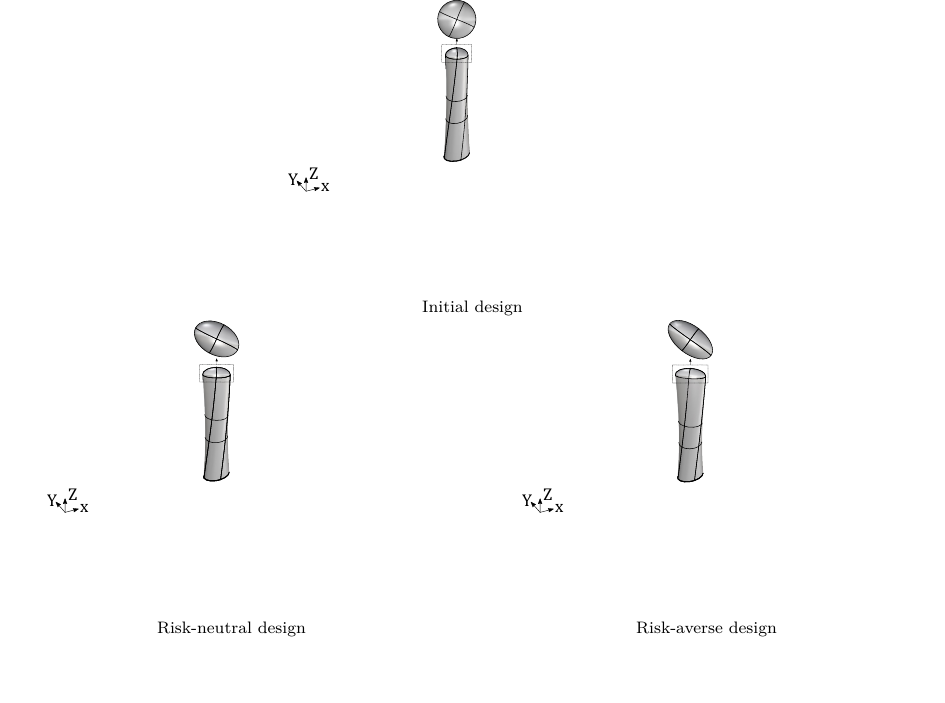}
    \caption{Numerical study II: initial and final building designs.}
    \label{fig:result_all_two}
\end{figure}

\begin{figure}
    \centering
    \includegraphics[width=0.9\textwidth]{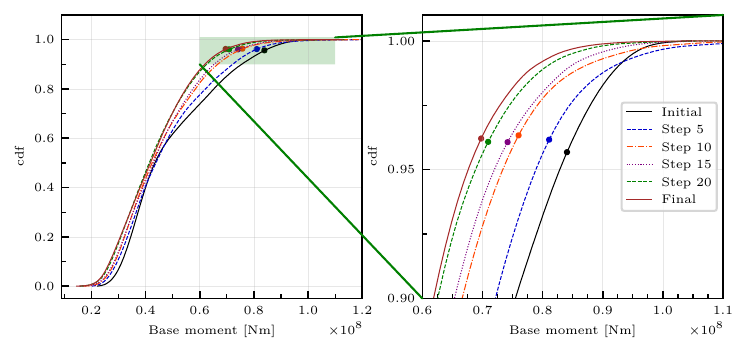}
\caption{Numerical study II: CDF of the base moment throughout the risk-averse optimization process. $\mathrm{CVaR}_{0.90}(M_{\mathcal{Z}})$ is represented as a dot in the cdf in the zoomed plot on the right.}
\label{fig:result_cdf}
\end{figure}

\begin{figure}
	\centering
	\includegraphics[trim=0cm 4.5cm 0cm 0cm, clip=true,height=0.275\textheight]{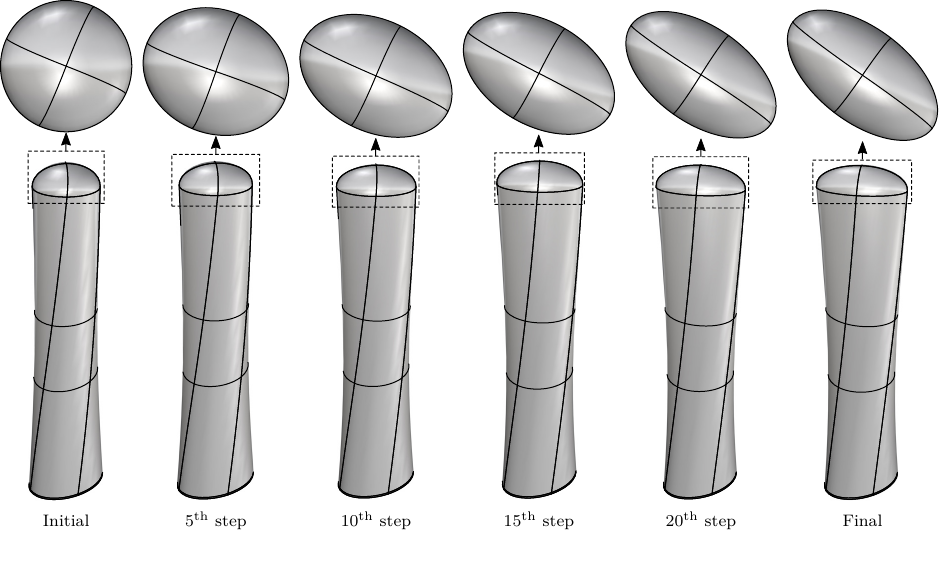}
    \caption{Numerical study II: shape change for risk-averse base moment optimization \cref{eq:prob2}.}
    \label{fig:result_shape_CVaR}
\end{figure}

The optimization record for the risk-averse candidate design is given in~\Cref{fig:result_cvar}.
Here, the improvement in the objective function $J(\mathcal{Z})$ is witnessed to be $\sim 17\%$. The evolution of the design parameters are shown in~\Cref{fig:result_cvar}.
We note that the number of samples required by the adaptive sampling algorithm increases as the design converges to the optimum. Compared to the cost of performing each iteration with the maximum number of samples, the adaptive sampling approach is 62.8\% cheaper.
\Cref{fig:result_shape_CVaR} shows the evolution of the geometry throughout the optimization process and \Cref{fig:result_cdf} shows the corresponding evolution of the CDF.
In \Cref{fig:result_shape_CVaR}, the tapering of the top cross-section is very prominent.
As in the risk-neutral setting, this may be the major contributing factor for the reduction in the objective function. The twist of the building is much more dramatic than in the risk-neutral setting.

\begin{figure}
    \centering
    \includegraphics[width=0.9\textwidth]{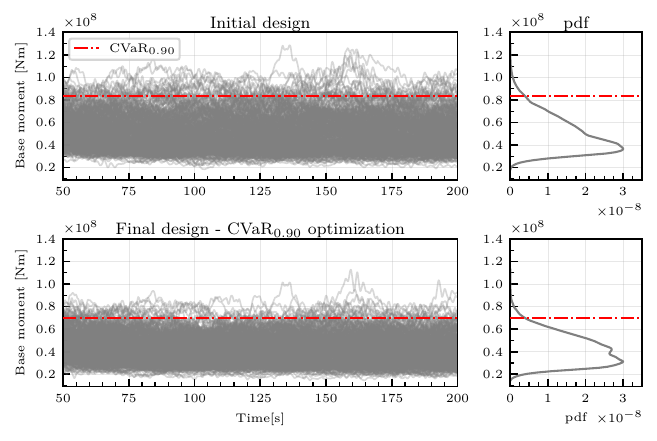}
\caption{Numerical study II: time series and PDFs of the base moment for \ref{eq:prob2}.}
\label{fig:result_timeseries}
\end{figure}

The time series and PDFs of the base moments for are shown in \Cref{fig:result_timeseries}.
As in Study I, we see a great improvement in the base moment distribution after shape optimization.
The final risk-neutral and risk-averse building designs are collected together in~\Cref{fig:result_all_two}.

\section{Conclusions} \label{sec:conclusions}
A workflow is developed and presented for stochastic optimization considering the uncertainties in the incoming wind for building design. Both risk-neutral and risk-averse optimization are compared. The approaches are illustrated by two numerical examples, and the performance in reducing the objective function is demonstrated. The adopted adaptive sampling is found to be effective in reducing the overall computational cost. For the first optimization problem, it is clearly witnessed that the optimal designs under uncertainty and the optimal design subject only to the predominant wind direction (PWD) are quite different. The PWD design is found to underperform when uncertainties are re-introduced compared to the designs that were optimized with uncertainties seen during the optimization process. This demonstrates the need for tall building design under uncertain wind conditions.
We then see that as the number of design parameters increases from one to two, risk-neutral and risk-averse optimization result in different solutions, with risk-averse optimization leading to a lower probability of high-loading scenarios. We demonstrate the benefits of CVaR (risk-averse) optimization over mean value (neutral) optimization for building design, in particular to control extreme values of loads that are of significant consequence to the safety of the final built structure.

\section*{Acknowledgments} \label{sec:acknowledgments}

We thank David~Andersson, Michael~Andre, Quentin~Ayoul-Guilmard, Dagmawi~Bekel, Sami~Bidier, Sundar~Ganesh, Matthew~Keller, Alex~Michalski, Fabio~Nobile, Marc~Nu\~{n}ez, Riccardo~Rossi, Riccardo~Tosi, and the rest of the ExaQUte project for helpful discussions.
This project has received funding from the European Union's Horizon 2020 research and innovation programme under grant agreement No 800898.
UK and BW gratefully acknowledge the support of the Deutsche Forschungsgemeinschaft (DFG) within the project WO 671/11-1. AK acknowledges the support of Deutsche Akademische Austauschdienst (DAAD).

\section*{Disclaimer} \label{sec:disclaimer}

This work was performed under the auspices of the U.S. Department of Energy by Lawrence Livermore National Laboratory under Contract DE--AC52--07NA27344 and the LLNL-LDRD Program under Project tracking No.\ 22--ERD--009.
Release number LLNL-JRNL-832223.

\appendix
\section{Copula-based model for wind speed and direction} \label{sec:copula_based_model_for_wind_speed_and_direction}

In this appendix, we describe the construction of our copula-based statistical model for the bivariate distribution of $\theta$ and $\overline{u}$.\footnote{Recall from \cref{foot:ubar} that a statistical model for $\theta$ and $\overline{u}$, at a fixed height $z$, can also be used as a statistical model for $\theta$ and $u_\ast$.}
A bivariate copula $C(\cdot,\cdot)$ is a bivariate distribution on the unit square $[0,1]^2$ with uniform $\mathrm{Unif}(0,1)$ marginals.
Multivariate copulas are often used to construct low-dimensional statistical models for random variables with complicated dependencies.

\begin{figure}
\centering
    \includegraphics[width=\textwidth]{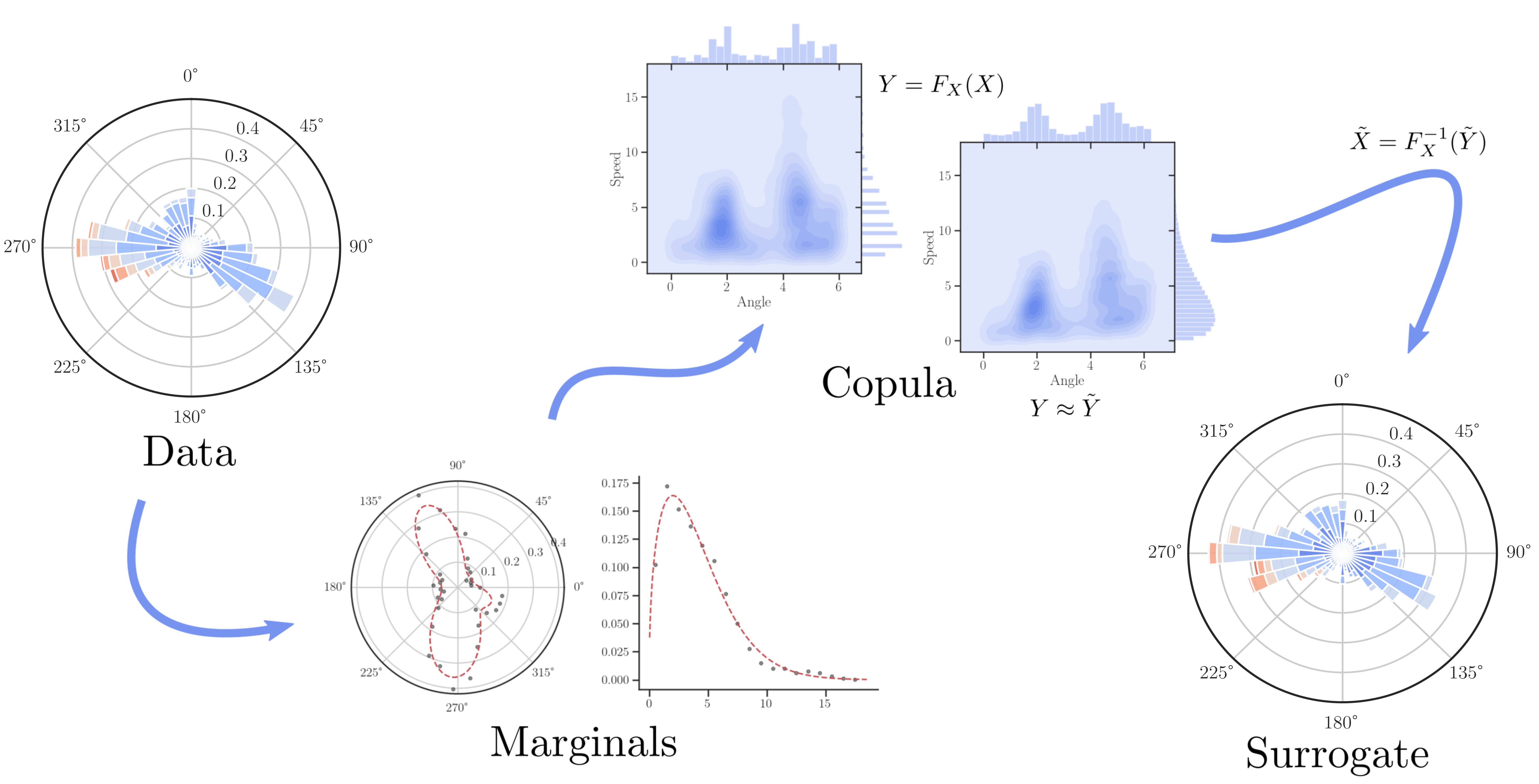}
    \caption{\label{fig:CopulaDiag} Construction of the bivariate joint distribution for $\overline{u}$ and $\theta$.}
\end{figure}

Let $X$ and $Y$ be random variables.
Sklar's theorem \cite{sklar1959fonctions} states that for every joint distribution $F_{X,Y}(\cdot,\cdot)$, there exists a copula $C(\cdot,\cdot)$ such that
\begin{equation}
	F_{X,Y}(x,y) = C(F_X(x),F_Y(y))
	\,,
\end{equation}
where $F_X(\cdot)$ and $F_Y(\cdot)$ are the marginals of $F_{X,Y}(\cdot,\cdot)$.
A wide variety of copula models exist in the literature \cite{nelsen2007introduction}.
In this work, we use an empirical copula, $\hat{C}_n(\cdot,\cdot)$, described below and illustrated in \Cref{fig:CopulaDiag}.

Let $F_{\theta,\bar{u}}(x,y)$ be the bivariate distribution function for the wind angle and mean velocity at 10m above ground.
Let $\{(\theta_i,\bar{u}_i)\}_{i=1}^n$ be a set of samples from the the joint distribution of $(\theta,\bar{u})$.
Next, define $\{(u_i,v_i)\}_{i=1}^n$ to be the transformed samples $(u_i,v_i) = (\hat{F}_\theta(\theta_i),\hat{F}_{\bar{u}}(\bar{u}_i))$, where $\hat{F}_\theta \approx F_\theta$ and $\hat{F}_{\bar{u}} \approx F_{\bar{u}}$ are consistent estimators of the true marginals.
In this work, we constructed $\hat{F}_\theta$ using a maximum likelihood estimate of a mixture of von Mises distributions with a prescribed orientation, and $\hat{F}_{\bar{u}}$ was constructed from the maximum likelihood estimate of a Weibull distribution.\footnote{For an illustration, see the two plots with the subtitle "Marginals" in \Cref{fig:CopulaDiag}.}
Finally, the empirical copula can be defined from the set $\{(u_i,v_i)\}_{i=1}^n$ as follows:
\begin{equation}
	\hat{C}_n(u,v)
	=
	\frac{1}{n}\sum_{i=1}^n I(u_i\leq u, v_i \leq v)
	\,,
\end{equation}
where $I(a\leq b, c\leq d)$ is the indicator function
\begin{equation}
	I(a\leq b, c\leq d)
	=
	\begin{cases}
		1 &\text{if } a\leq b,~ c\leq d, \\
		0 &\text{otherwise.}
	\end{cases}
\end{equation}
The set of transformed samples is in the unit square $[0,1]^2$.

With these definitions in hand, we now define the empirical, copula-based distribution function
\begin{equation}
	\hat{F}_{(\theta,\bar{u})}(\theta,\bar{u})
	=
	\hat{C}_n(\hat{F}_{\theta}(\theta),\hat{F}_{\bar{u}}(\bar{u}))
	\,.
\end{equation}
This function can be used to draw dependent samples of $\theta$ and $\bar{u}$.
For illustration, $2500$ samples were drawn to create the wind rose on the right-hand side of \Cref{fig:WindRoses}.

\phantomsection\bibliographystyle{abbrv}
\bibliography{main}

\end{document}